\documentclass[twocolumn]{aastex62}

\usepackage[T1]{fontenc}
\usepackage[utf8]{inputenc}
\usepackage{amsmath}

\usepackage{tikz}
\usetikzlibrary{shapes, arrows, positioning, calc}

\tikzset{
    base/.style = {text width=3cm, text centered,},
    node/.style = {base, draw, minimum height=1.5cm, fill=white, semithick, scale=0.8},
    process/.style = {node, rectangle, fill=black!15},
    data/.style = {node, rectangle, rounded corners},
    line/.style = {draw, semithick, color=black!90},
    arrow/.style = {line, -to},
    arrow-dashed/.style = {arrow, dashed},
    -|-/.style={
        to path={
            (\tikztostart) -| ($(\tikztostart)!#1!(\tikztotarget)$) |- (\tikztotarget)
            \tikztonodes
        }
    },
    -|-/.default=0.5,
    |-|/.style={
        to path={
            (\tikztostart) |- ($(\tikztostart)!#1!(\tikztotarget)$) -| (\tikztotarget)
            \tikztonodes
        }
    },
    |-|/.default=0.5,
}

\newcommand{\revision}[1]{#1}

\renewcommand{\b}{\pmb}
\renewcommand{\d}{{\rm d}}
\DeclareMathOperator{\tr}{tr} 

\newcommand{\astrocaltech}{
    \affiliation{
        Department of Astronomy,
        California Institute of Technology,
        1200 E California Blvd,
        Pasadena, CA 91125, USA
    }
}

\newcommand{\eecaltech}{
    \affiliation{
        Department of Electrical Engineering,
        California Institute of Technology,
        1200 E California Blvd,
        Pasadena, CA 91125, USA
    }
}

\newcommand{\ovro}{
    \affiliation{
        California Institute of Technology,
        Owens Valley Radio Observatory,
        Big Pine, CA 93513, USA
    }
}

\newcommand{\harvard}{
    \affiliation{
        Harvard-Smithsonian Center for Astrophysics,
        60 Garden Street,
        Cambridge, MA 02138, USA
    }
}

\newcommand{\swinburne}{
    \affiliation{
        Centre for Astrophysics \& Supercomputing,
        Swinburne University of Technology,
        Hawthorn, VIC 3122, Australia
    }
}

\newcommand{\berkeley}{
    \affiliation{
        Department of Astronomy,
        University of California Berkeley,
        Berkeley, CA 94720, USA
    }
}

\newcommand{\virtualitics}{
    \affiliation{
        Virtualitics,
        225 S Lake Ave,
        Pasadena, CA 91101
    }
}

\newcommand{\astron}{
    \affiliation{
        Netherlands Institute for Radio Astronomy (ASTRON),
        Oude Hoogeveensedijk 4, 7991PD,
        Dwingeloo, The Netherlands
    }
}

\newcommand{\unm}{
    \affiliation{
        Department of Physics and Astronomy,
        University of New Mexico,
        Albuquerque, NM 87131, USA
    }
}

\newcommand{\nrao}{
    \affiliation{
        National Radio Astronomy Observatory,
        P.O. Box O,
        Socorro, NM 87801, USA
    }
}

\newcommand{\nrl}{
    \affiliation{
        Naval Research Laboratory,
        Code 7213,
        Washington, DC 20375, USA
    }
}

\begin{document}


\title{
    The 21\,cm Power Spectrum from the Cosmic Dawn: First Results from the OVRO-LWA
}

\author[0000-0002-4731-6083]{Michael~W.~Eastwood}  \astrocaltech
\author{Marin~M.~Anderson}    \astrocaltech
\author{Ryan~M.~Monroe}       \eecaltech
\author{Gregg~Hallinan}       \astrocaltech
\author{Morgan~Catha}         \ovro
\author[0000-0003-1407-0141]{Jayce~Dowell}   \unm
\author{Hugh~Garsden}         \harvard
\author{Lincoln~J.~Greenhill} \harvard
\author{Brian~C.~Hicks}       \nrl
\author{Jonathon~Kocz}        \astrocaltech
\author[0000-0003-2783-1608]{Danny~C.~Price} \swinburne \berkeley \harvard
\author{Frank~K.~Schinzel}    \nrao
\author{Harish~Vedantham}     \astrocaltech \astron
\author{Yuankun~Wang}         \astrocaltech \virtualitics

\correspondingauthor{Michael W. Eastwood}
\email{mweastwood@astro.caltech.edu}

\begin{abstract}
    The 21\,cm transition of neutral hydrogen is opening an observational window into the cosmic
    dawn of the universe---the epoch of first star formation. We use 28\,hr of data from the Owens
    Valley Radio Observatory Long Wavelength Array (OVRO-LWA) to place upper limits on the spatial
    power spectrum of 21\,cm emission at $z \approx 18.4$ ($\Delta_{21} \lesssim 10^4\,\text{mK}$),
    and within the absorption feature reported by the EDGES experiment \citep{2018Natur.555...67B}.
    In the process we demonstrate the first application of the double Karhunen-Lo\`{e}ve transform
    for foreground filtering, and diagnose the systematic errors that are currently limiting the
    measurement. We also provide an updated model for the angular power spectrum of low-frequency
    foreground emission measured from the northern hemisphere, which can be used to refine
    sensitivity forecasts for next-generation experiments.
\end{abstract}


\section{Introduction}\label{sec:introduction}

The Cosmic Dawn of star formation is one of the final unexplored epochs of the universe. During this
time (approximately $25 \gtrsim z \gtrsim 15$) the first generation of stars and galaxies formed and
brought an end to the Dark Ages. Ly$\alpha$ emission from this early star formation couples the
excitation temperature of the 21\,cm hyperfine structure transition (i.e., the spin temperature) to
the local gas temperature of the intergalactic medium \citep[IGM;][]{1952AJ.....57R..31W,
1958PIRE...46..240F}. This allows the highly-redshifted 21\,cm transition to be used as a probe of
the density, temperature, and ionization state of the IGM \citep[e.g.,][]{2006PhR...433..181F,
2012RPPh...75h6901P}.

The first possible detection of high-redshift ($z\sim 17$) atomic hydrogen in the globally averaged
sky temperature was recently reported by the EDGES experiment at 78\,MHz
\citep{2018Natur.555...67B}.  This measurement was remarkable for its extreme amplitude ($\sim
500\,\text{mK}$), and unusual width and shape.  A plethora of new ideas have been proposed to
explain the amplitude of the absorption trough. These new theories generally fall into two
categories: those that invoke new physics to cool the IGM at a rate faster than pure adiabatic
cooling would otherwise allow \citep[e.g.,][]{2018Natur.555...71B, 2018PhRvL.121a1101F}, and those
that posit a new radio background originating from $z\gtrsim 20$ \citep[e.g.][]{2018arXiv180301815E,
2018ApJ...858L...9D}.  Distinguishing between these alternatives, and confirming the existing
measurement \citep[e.g., some concerns about the detection have been raised
by][]{2018arXiv180501421H} now motivates a number of experiments.

The LEDA \citep{2018MNRAS.478.4193P} and SARAS~2 \citep{2018ApJ...858...54S} experiments are in the
process of attempting to directly confirm the EDGES detection in the global sky temperature using
radiometric dipole antennas.  Notably, each of these experiments employ materially different antenna
designs, which will, in principle, help address concerns regarding the role of the antenna beam and
its potential to introduce spectral structure into the measurement. As noted by
\citet{2018Natur.555...67B}, these independent measurements with independent processing pipelines
will be an important verification of an exceptionally difficult measurement.  The primary observing
challenge faced by global-detection experiments---such as EDGES, LEDA, and SARAS~2---is controlling
systematic errors introduced by foreground radio emission, instrumental effects, and the interaction
between them. Although these experiments calibrate their antennas and electronics with great care,
they must rely on external models of the foreground radio emission to model many types of systematic
errors.

In contrast, interferometers generally have the ability to self-calibrate and build self-consistent
models for the sky emission \citep{1978ApJ...223...25R}, even with the extremely wide fields of view
common for low-frequency interferometers \citep[e.g.,][]{2018AJ....156...32E}. However,
interferometers are generally not used to measure the globally averaged sky brightness
\citep{2016ApJ...826..116V}, but instead measure the three dimensional spatial power spectrum of the
21\,cm brightness temperature fluctuations. The global average and the spatial power spectrum are
both statistics of the same field and therefore a measurement of the spatial power spectrum can also
provide evidence to support a reject a putative detection in the global average.  A detection of the
spatial power spectrum will---with some modeling---provide independent constraints on the
temperature of the IGM and the timing of early star and galaxy formation
\citep[e.g.,][]{2017MNRAS.472.2651G}.  The spatial power spectrum carries additional information
about the scale of the brightness temperature fluctuations, which may be used to constrain, for
example, the amplitude of Lyman--Werner feedback \citep{2013MNRAS.432.2909F} and the spectral
hardness of early X-ray sources \citep{2014MNRAS.437L..36F} that heat the IGM.

At lower redshifts corresponding to the Epoch of Reionization (EoR), constraints on the 21\,cm
spatial power spectrum have been published by the PAPER experiment \citep{2015ApJ...809...61A},
LOFAR \citep{2017ApJ...838...65P}, the MWA \citep{2016ApJ...833..102B}, and the GMRT
\citep{2013MNRAS.433..639P}. At redshifts corresponding to the Cosmic Dawn, this measurement has
only been previously attempted by \citet{2016MNRAS.460.4320E} using 6\,hr of data from the MWA.  The
under-construction HERA experiment will aim to place the most sensitive limits to date on the 21\,cm
brightness temperature spatial power spectrum from both the EoR and Cosmic Dawn due to its large
collecting area and design lessons inherited from the PAPER experiment \citep{2017PASP..129d5001D}.
Similarly, the SKA will build on the development of its pathfinder arrays to measure the spatial
power spectrum and image the three-dimensional structure of the universe through the 21\,cm
transition \citep{2013ExA....36..235M, 2015aska.confE...1K}.

In this paper we present the first attempted measurement of the spatial power spectrum of 21\,cm
brightness temperature fluctuations with the OVRO-LWA. In the process, we model, derive, and (where
appropriate) measure the contribution of thermal noise, foreground emission, and the 21\,cm signal
to the full covariance matrix of the data. This is possible due to the application of $m$-mode
analysis \citep{2014ApJ...781...57S, 2015PhRvD..91h3514S}, which introduces sparsity into the
covariance matrices without which it would not be possible to store the full covariance matrix of
the data. \revision{
    We also analyze the systematic sources of error that currently limit the measurement through
    simulation, which allows us to derive quantitative requirements for instrumental calibration
    errors.
}

In \S\ref{sec:observations} we describe the observations, the calibration strategy, and point source
removal routines used in this work. In \S\ref{sec:formalism} we will describe the $m$-mode analysis
formalism and a new strategy for compressing the representation of the transfer matrix.  In
\S\ref{sec:sensitivity} we derive, model, and measure the contribution of noise, foreground emission
and the cosmological 21\,cm signal to the full covariance matrix of the measured data.  These
covariance matrices are applied to filter the foreground emission in
\S\ref{sec:foreground-filtering}, where we also build physical intuition for the action of the
foreground filters derived by \citep{2014ApJ...781...57S, 2015PhRvD..91h3514S}.  These foreground
filters are applied to 28\,hr of data from the OVRO-LWA to estimate the 21\,cm power spectrum in
\S\ref{sec:results}, where we also analyze the limiting systematic errors in our measurement.
Finally, in \S\ref{sec:conclusion} we present our conclusions.  Unless stated otherwise, we adopt
the set of cosmological parameters measured by \citet{2016A&A...594A..13P}.

\section{Observations}\label{sec:observations}

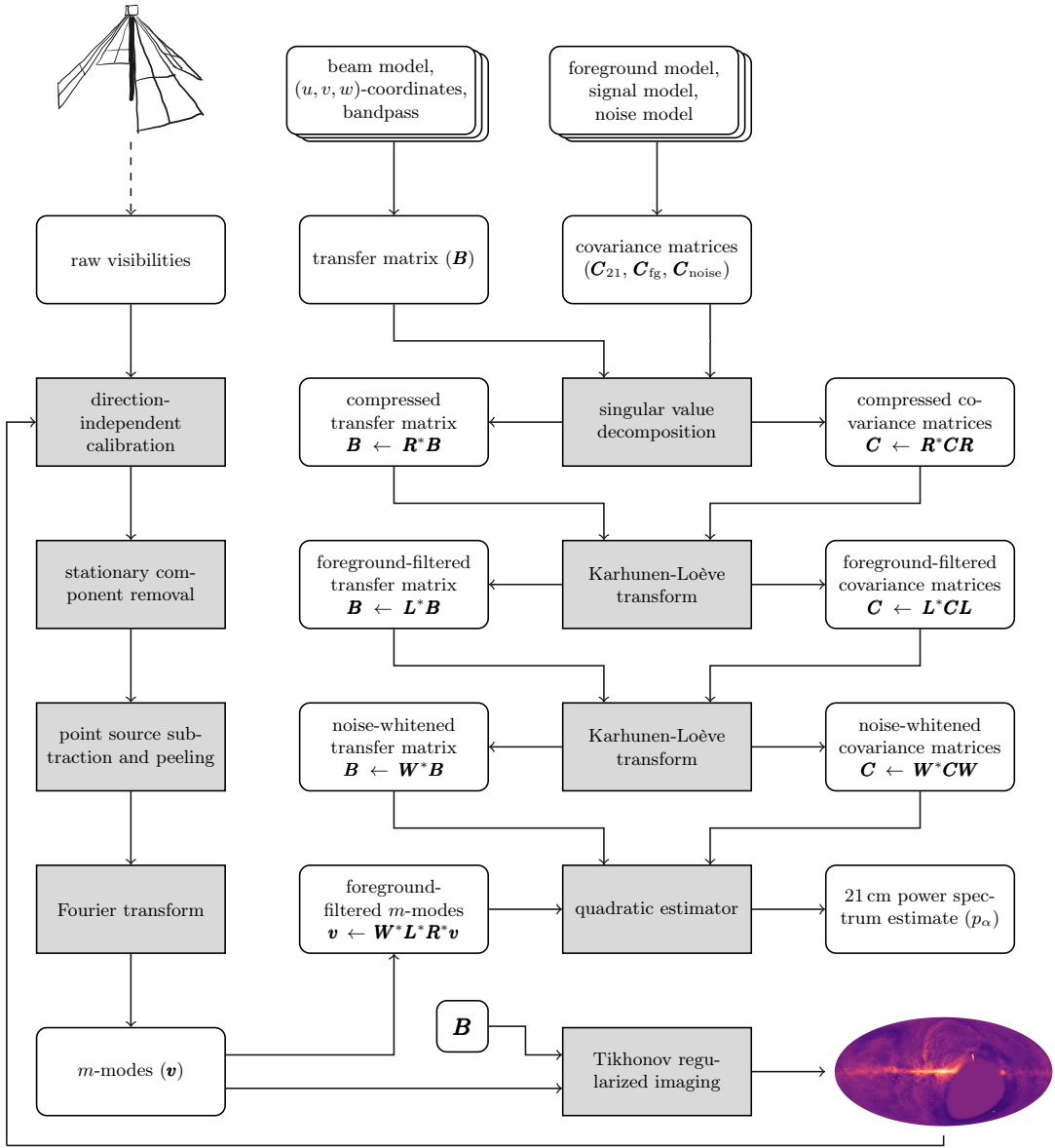
\begin{figure*}
    \centering
    \begin{tikzpicture}[node distance=1cm]

        \node [base] (antenna) {\includegraphics[width=2cm]{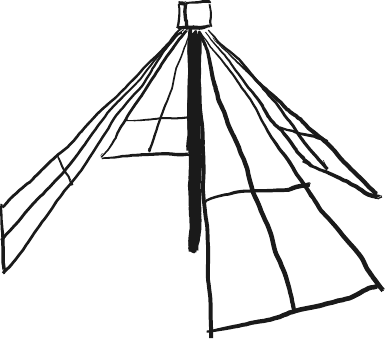}};
        \node [data, below=of antenna]       (raw)        {raw visibilities};
        \node [process, below=of raw]        (calibrate)  {direction-independent calibration};
        \node [process, below=of calibrate]  (stationary) {stationary component removal};
        \node [process, below=of stationary] (peeling)    {point source subtraction and peeling};
        \node [process, below=of peeling]    (fourier)    {Fourier transform};
        \node [data, below=of fourier]       (mmodes)     {$m$-modes ($\b v$)};

        \node [data, right=of raw]      (transfer) {transfer matrix ($\b B$)};

        \node [data, above=of transfer] (beam1) {};
        \node [data] (beam2) at ($ (beam1) + (-1cm:-0.1cm) $) {};
        \node [data] (beam3) at ($ (beam2) + (-1cm:-0.1cm) $) {
            beam model, $(u, v, w)$-coordinates, bandpass
        };

        \node [data, right=of beam1] (sky-model1) {};
        \node [data] (sky-model2) at ($ (sky-model1) + (-1cm:-0.1cm) $) {};
        \node [data] (sky-model3) at ($ (sky-model2) + (-1cm:-0.1cm) $) {
            foreground model,\\signal model,\\noise model
        };

        \node [data, below=of sky-model1] (covariance)
            {covariance matrices ($\b C_\text{21}$, $\b C_\text{fg}$, $\b C_\text{noise}$)};

        \node [data, below=of transfer] (compressed-transfer)
            {compressed transfer matrix\\$\b B\leftarrow\b R^*\b B$};
        \node [process, right=of compressed-transfer] (svd) {singular value decomposition};
        \node [data, right=of svd] (compressed-covariance)
            {compressed covariance matrices\\$\b C\leftarrow\b R^*\b C\b R$};

        \node [data, below=of compressed-transfer] (filtered-transfer)
            {foreground-filtered transfer matrix\\$\b B\leftarrow\b L^*\b B$};
        \node [process, right=of filtered-transfer] (kltransform1) {Karhunen-Lo\`{e}ve transform};
        \node [data, right=of kltransform1] (filtered-covariance)
            {foreground-filtered covariance matrices\\$\b C\leftarrow\b L^*\b C\b L$};

        \node [data, below=of filtered-transfer] (whitened-transfer)
            {noise-whitened transfer matrix\\$\b B\leftarrow\b W^*\b B$};
        \node [process, right=of whitened-transfer] (kltransform2) {Karhunen-Lo\`{e}ve transform};
        \node [data, right=of kltransform2] (whitened-covariance)
            {noise-whitened covariance matrices\\$\b C\leftarrow\b W^*\b C\b W$};

        \node [process, below=of kltransform2] (qestimator) {quadratic estimator};
        \node [data, left=of qestimator] (filtered-mmodes)
            {foreground-filtered $m$-modes\\$\b v\leftarrow\b W^*\b L^*\b R^*\b v$};
        \node [data, right=of qestimator] (ps)
            {21\,cm power spectrum estimate ($p_\alpha$)};

        \node [process, below=of qestimator] (tikhonov)
            {Tikhonov regularized imaging};
        \node [base, right=of tikhonov] (maps)
            {
                \includegraphics[width=3cm]{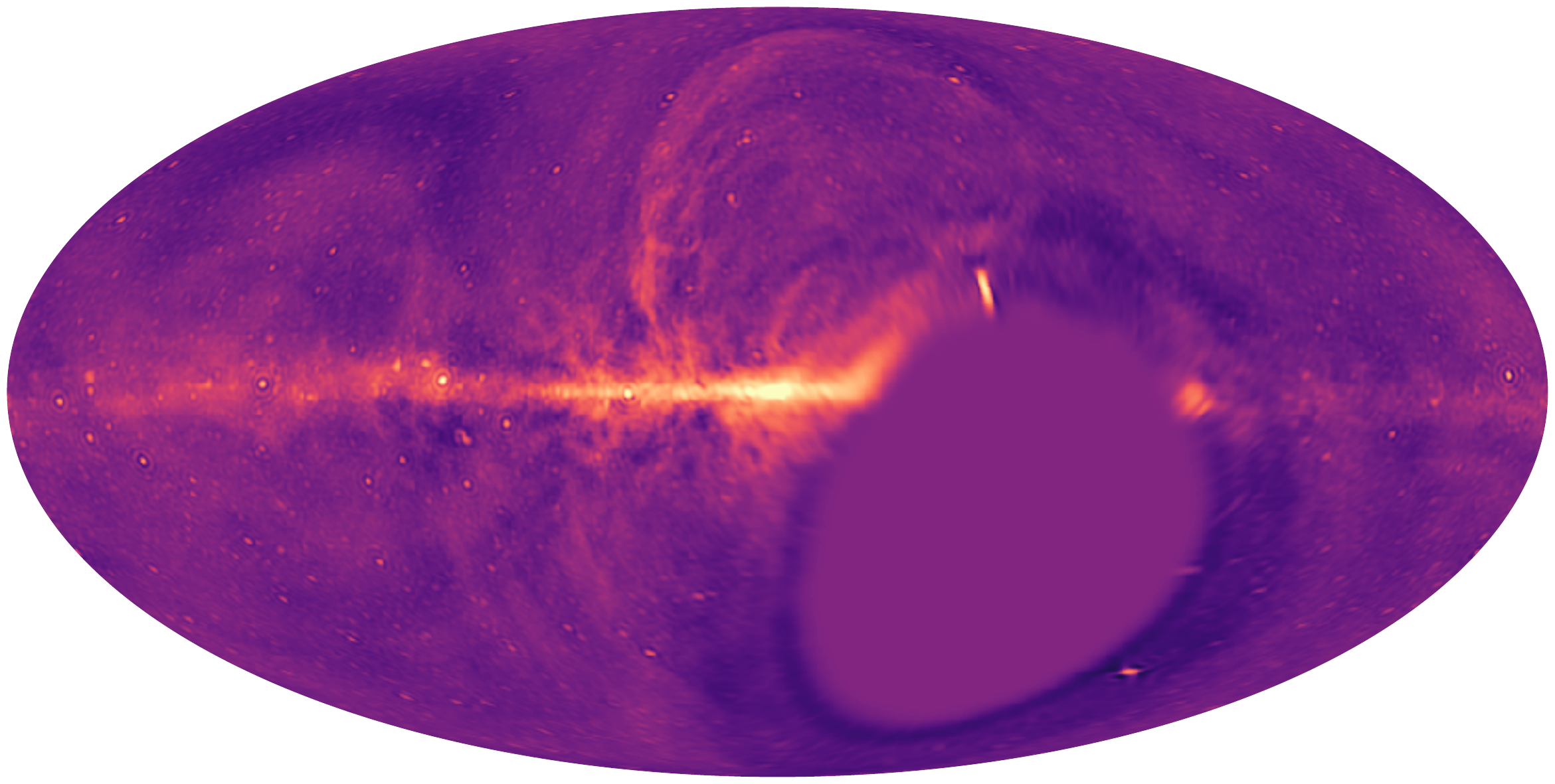}
            };

        \node [rectangle, above left=-3.5mm and 1cm of tikhonov, rounded corners,
               draw, semithick, minimum width=7mm, minimum height=7mm] (B) {$\b B$};

        \coordinate [below left=+4mm and 4mm of mmodes] (corner);


        \path [arrow-dashed] (antenna) -- (raw);
        \path [arrow] (raw) -- (calibrate);
        \path [arrow] (calibrate) -- (stationary);
        \path [arrow] (stationary) -- (peeling);
        \path [arrow] (peeling) -- (fourier);
        \path [arrow] (fourier) -- (mmodes);

        \path [arrow] (beam1) -- (transfer);
        \path [arrow] (transfer.south) to[|-|] (svd.140);
        \path [arrow] (svd) -- (compressed-transfer);
        \path [arrow] (compressed-transfer.south) to[|-|] (kltransform1.140);
        \path [arrow] (kltransform1) -- (filtered-transfer);
        \path [arrow] (filtered-transfer.south) to[|-|] (kltransform2.140);
        \path [arrow] (kltransform2) -- (whitened-transfer);

        \path [arrow] (sky-model1) -- (covariance);
        \path [arrow] (covariance.320) to[|-|] (svd.40);
        \path [arrow] (svd) -- (compressed-covariance);
        \path [arrow] (compressed-covariance.south) to[|-|] (kltransform1.40);
        \path [arrow] (kltransform1) -- (filtered-covariance);
        \path [arrow] (filtered-covariance.south) to[|-|] (kltransform2.40);
        \path [arrow] (kltransform2) -- (whitened-covariance);

        \path [arrow] (whitened-transfer.south) to[|-|] (qestimator.140);
        \path [arrow] (whitened-covariance.south) to[|-|] (qestimator.40);
        \path [arrow] (mmodes.10) -| (filtered-mmodes.south);
        \path [arrow] (filtered-mmodes) -- (qestimator);
        \path [arrow] (qestimator) -- (ps);

        \path [arrow] (mmodes.350) -- (tikhonov.190);
        \path [arrow] (B.east) to[-|-] (tikhonov.170);
        \path [arrow] (tikhonov) -- (maps);

        \path [line] (maps) |- (corner);
        \path [arrow] (corner) |- (calibrate);

    \end{tikzpicture}
    \caption{
        A flow chart describing the data analysis steps performed in this paper.  \revision{Shaded
        boxes represent processing steps, whereas unshaded boxes represent data.} Radio waves are
        received by antennas (depicted in the upper-left corner), which are correlated to produce
        raw visibilities. These visibilities are then flagged, calibrated, and bright point sources
        are removed. After a full sidereal day's worth of data has been collected, these
        visibilities can be Fourier transformed to compute the measured $m$-modes.  Separately, an
        empirical beam model is used to calculate the transfer matrix elements that describe the
        interferometer's sensitivity to the sky. Full covariance matrices are computed for the
        foreground emission, 21\,cm signal, and thermal noise. These matrices are used to compress,
        filter foreground emission, and whiten the noise covariance. Finally, the resulting filtered
        $m$-modes are used to estimate the spatial power spectrum of 21\,cm emission. Images of the
        sky can be constructed through the use of Tikhonov-regularized imaging
        \citep{2018AJ....156...32E}, which are useful for diagnosing errors in the analysis.
    }
    \label{fig:flowchart}
\end{figure*}

We collected 28\,hr of continuous data using the OVRO-LWA beginning at 2017 February 17 12:00:00
UTC. The OVRO-LWA is a low-frequency radio interferometer with a bandpass covering 27--85\,MHz ($50
\gtrsim z \gtrsim 16$), and is currently composed of 288 dual-polarization dipole antennas (Hallinan
et al., in prep.).  251 of these antennas are arranged within a dense 200\,m diameter core in a
configuration optimized for sidelobe levels in snapshot images.  32 additional expansion antennas
are placed outside of the core, expanding the maximum baseline length to 1.5\,km. The remaining five
antennas are equipped with radiometric front-ends for total power measurements of the sky as part of
the Large-Aperture Experiment to Detect the Dark Ages \citep[LEDA;][]{2018MNRAS.478.4193P}.  The
LEDA correlator \citep{2015JAI.....450003K} serves as the back-end for the OVRO-LWA, and
cross-correlates 512 inputs with 58\,MHz instantaneous bandwidth.  In this configuration the
OVRO-LWA performs full cross-correlation of 256 antennas (512 signal paths), and 32 antennas (64
signal paths) are unused.  We selected the correlator's integration time to be 13\,s as this evenly
divides the sidereal day to within 0.1\,s.  In snapshot images, the OVRO-LWA can capture the entire
visible hemisphere at $10\arcmin$ resolution \citep[e.g.,][]{2017arXiv171106665A}, and this same
dataset was used to generate maps of the sky north of $\delta=-30\arcdeg$
\citep{2018AJ....156...32E}.

At low radio frequencies, propagation effects through the ionosphere are important.  During this
observing period, however, geomagnetic and ionospheric conditions were mild. At 73\,MHz, bright
point sources were observed to refract by up to $4\arcmin$ on 10\,minute timescales, and the
apparent flux of point sources varied by up to 10\% on 13\,s timescales (averaging over 24\,kHz
bandwidth) due to ionospheric conditions.

In this work we selected data from an instrumental subband centered at 73.152\,MHz with 2.6\,MHz
bandwidth ($z=18.4$, $\Delta z=0.8$). This subband is contained within the absorption feature
observed by \citet{2018Natur.555...67B}, and contains the 73.0--74.6\,MHz band allocated for radio
astronomy in the United States. There is additionally a gap in television broadcasting between
72\,MHz (the upper edge of channel 4) and 76\,MHz (the lower edge of channel 5) that this observing
band takes advantage of.  Additionally, in previous work we published an updated low-frequency sky
map at 73.152\,MHz \citep{2018AJ....156...32E} that is available online at the Legacy Archive for
Microwave Background Data Analysis (LAMBDA).

When measuring the power spectrum of 21\,cm fluctuations, it is common to make an implicit
assumption that the 21\,cm power spectrum is not evolving along the line-of-sight direction (see
Appendix~\ref{app:spatial-to-angular}).  \citet{2018MNRAS.477.3217G} simulated this effect and found
that for volumes of equal comoving radial distance, this light-cone effect is more severe during the
Cosmic Dawn than during the EoR.  Near $z \sim 18$ and a volume with $\Delta z \sim 3$, the
recovered spatial power spectrum is suppressed by a factor $\lesssim 2$. While the light-cone effect
can limit the usable bandwidth for estimating the 21\,cm spatial power spectrum, we conclude that
2.6\,MHz of bandwidth is permissible for this initial analysis. Future studies of 21\,cm
fluctuations of the Cosmic Dawn, however, should instead consider estimating the multi-frequency
angular power spectrum \citep{2007MNRAS.378..119D}, which is a statistic that is less common in the
literature, but can be measured without assuming that the statistics of the fluctuations are not
evolving along the line of sight.

A summary of the analysis steps performed in this work---including the instrumental calibration and
21\,cm power spectrum reduction---can be seen in Figure~\ref{fig:flowchart}.  In particular, a gain
calibration was derived from a 45\,minute track of data beginning at 2017 February 17 17:46:28
during which the two brightest point sources in the northern hemisphere (Cyg~A and Cas~A) are near
the meridian. The sky model is initially composed of Cyg~A and Cas~A where the absolute spectrum of
Cyg~A is given by \citet{1977A&A....61...99B}, and the spectrum of Cas~A is adjusted for its secular
decrease of 0.77\% per year \citep{2009AJ....138..838H}. Because this initial sky model is
incomplete on large angular scales, baselines shorter than 15 wavelengths are excluded from the
calibration routine.  The gains are optimized using a variant of alternating least squares
independently described by \citet{2008ISTSP...2..707M} and \citet{2014A&A...571A..97S}.  The
bandpass amplitude is fit with a 5th order polynomial, and the phase is fit with a term for the
delay and a term for dispersion through the ionosphere. Smoothing the gain calibration in this way
helps to avoid modeling errors during calibration propagating into bandpass errors that can limit
the sensitivity of the interferometer to the 21\,cm power spectrum \citep{2016MNRAS.461.3135B,
2017MNRAS.470.1849E}. After this initial calibration and source removal, a model of the diffuse
galactic emission is constructed using Tikhonov-regularized $m$-mode analysis imaging. This model is
then used to recalibrate the data with a more complete model of the sky.

The OVRO-LWA analog signal path is susceptible to additive common-mode radio frequency interference
(RFI). A model for the common-mode RFI is constructed from the gain-calibrated visibilities after
averaging over the entire 28\,hr observing period with the phase center left at zenith. Averaging
the visibilities in this way smears out the contribution of the sky along characteristic sidereal
tracks. We then select the dominant components of the averaged visibilities to be used as templates
for the RFI. The templates are manually inspected for residual sky emission by imaging each
component with WSCLEAN \citep{2014MNRAS.444..606O}, and checking for features that are swept along
sidereal tracks. These templates are scaled and subtracted from each integration to suppress the
contamination of the common-mode RFI.

The top panel of Figure~\ref{fig:before-after-source-removal-sky-maps} is a dirty image of the sky
constructed from this dataset prior to any point source removal.  A handful of bright point sources
occupy the northern sky---namely Cas~A, Cyg~A, Her~A, Hya~A, Tau~A, Vir~A, 3C~123, 3C~353, and the
Sun.  Each of these sources is removed from the visibilities by employing a combination of peeling
(direction-dependent calibration) for the brightest sources, and source fitting and subtraction for
the fainter sources. This source removal strategy is described in greater detail by
\citet{2018AJ....156...32E}.

Finally, in order to reduce the data volume and computational cost of further reductions, we
selected only baselines representable with spherical harmonics with multipole number $l \le 300$.
This effectively selects only baselines from the core of the OVRO-LWA \revision{(23,947 baselines in
total)}, which contain the majority of the brightness temperature sensitivity. The data were
additionally averaged down to channel widths of 240\,kHz. At 73\,MHz, this averaging effectively
smears out the spatial power spectrum on $k_\parallel \approx 1\,\text{Mpc}^{-1}$ scales, but is
permissible because the expected cosmological signal is small on these scales.

\section{Formalism}\label{sec:formalism}

\subsection{$m$-Mode Analysis}

\begin{figure*}
    \centering
    \begin{tabular}{c}
        \includegraphics[width=\textwidth]{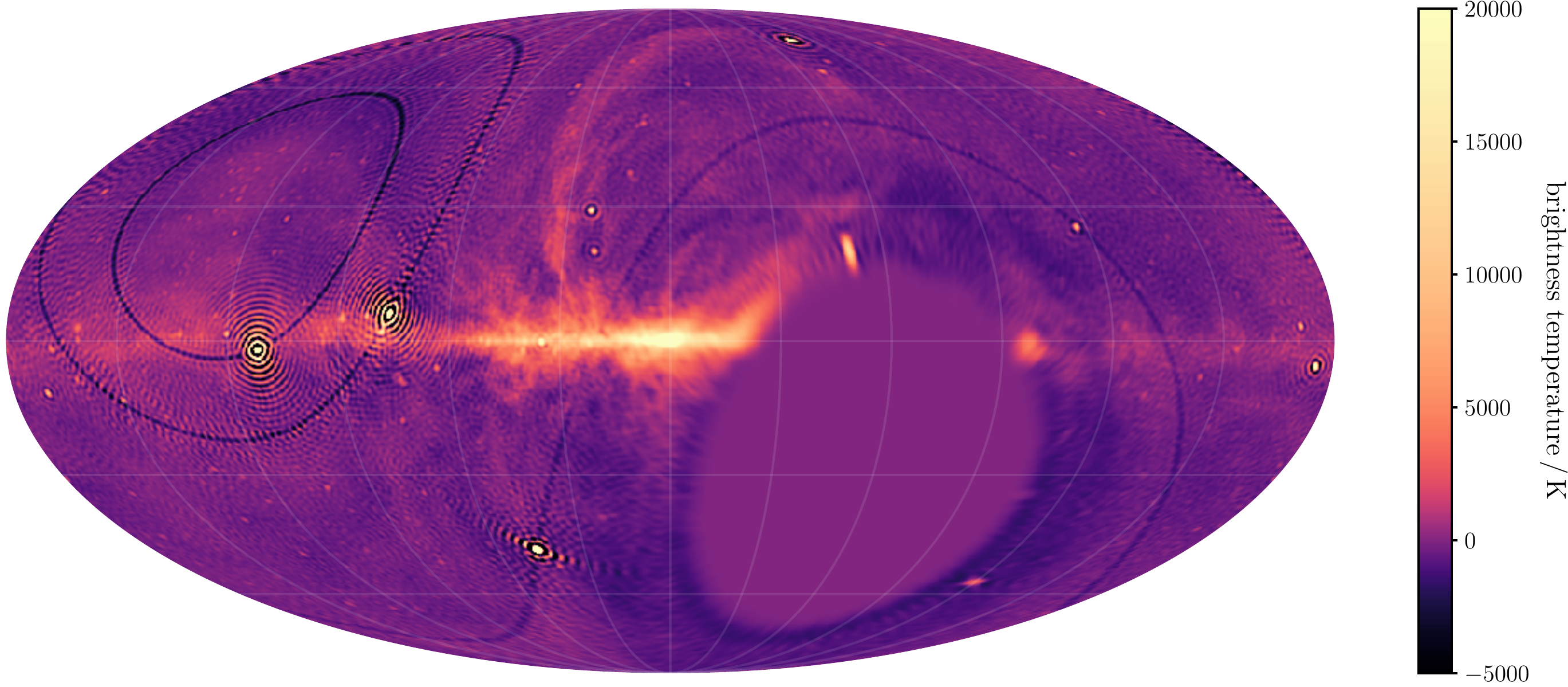} \\
        \includegraphics[width=\textwidth]{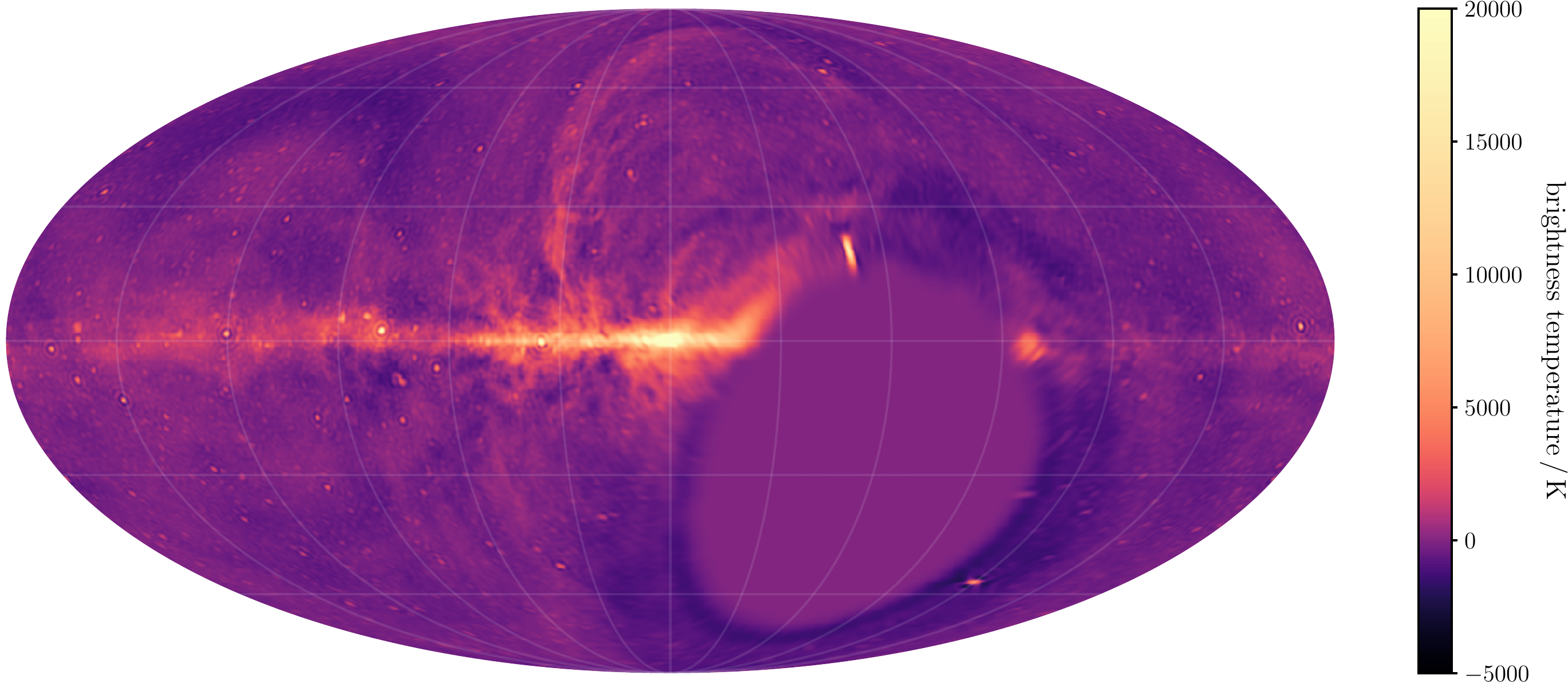} \\
    \end{tabular}
    \caption{
        A Molleweide projection of a Tikhonov-regularized image of the sky constructed from all
        baselines representable with $l_\text{max} \le 200$, and 2.6~MHz of bandwidth centered on
        73.2~MHz. The color scale is linear between $-1000$~K and $+1000$~K, and logarithmic outside
        of this range. No cleaning has been performed, so all point sources are convolved with a
        point spread function, and no masking of low declinations has been performed. The resolution
        of the maps naturally degrades at low declinations and the regularization scheme naturally
        encourages the map to be zero below the horizon. Negative rings at the declination of bright
        point sources are an artifact of the fact that $m=0$ modes are filtered from the dataset due
        to their susceptibility to RFI and common-mode pickup. (top) Before bright point sources are
        removed from the dataset. (bottom) After point source removal.
    }
    \label{fig:before-after-source-removal-sky-maps}
\end{figure*}

In this paper we apply the $m$-mode analysis formalism developed by \citet{2014ApJ...781...57S,
2015PhRvD..91h3514S}. The interested reader should consult the aforementioned references for
additional details, but $m$-mode analysis is briefly summarized below.

The measured quantity in a drift-scanning telescope is a periodic function of sidereal time.  The
Fourier transform with respect to sidereal time of this measured quantity is called an $m$-mode,
where the value of $m$ indicates how rapidly this mode varies over the course of a sidereal day.
$m=0$ corresponds to the mean value of the measurement over a sidereal day. $m=\pm1$ corresponds to
the components that vary once over a sidereal day. Larger absolute values of $m$ represent
contributions to the measurement that vary on increasingly rapid timescales.

The primary advantage of making this transformation to $m$-modes is that it can be shown that the
set of measured $m$-modes with a given value for $m$, are a linear combination of the spherical
harmonic coefficients with the same value of $m$. This allows the data to be partitioned by $m$, and
each partition can be manipulated independently of the remaining dataset. Typically this leads to a
large reduction in the processing time, which allows for the application of otherwise infeasible
data analysis techniques that make use of the full covariance matrix of the dataset.

We will adopt the convention that the measured $m$-modes are contained in a vector $\b v$, and the
spherical harmonic coefficients of the sky brightness are contained in a vector $\b a$. The transfer
matrix $\b B$ describes the interferometer's response to the sky and is block-diagonal when both $\b
v$ and $\b a$ are sorted by the absolute value of $m$. If we explicitly decompose the sky in terms
of the high-redshift 21\,cm contribution $\b a_\text{21}$, and the foreground radio emission $\b
a_\text{fg}$, then
\begin{equation}\label{eq:basic-m-mode-analysis}
    \b v = \b B \b a_\text{21} + \b B \b a_\text{fg} + \b n \,,
\end{equation}
where $\b n$ is the contribution of thermal noise to the measurement.

The rows of the transfer matrix $\b B$ fundamentally describe the response of each baseline to the
sky represented by $\b a$. The individual elements of the matrix are computed from spherical
harmonic transforms of each baseline's fringe pattern (including the response of the antenna beams
and bandpass). \citet{2018AJ....156...32E} demonstrated all-sky imaging in a single synthesis
imaging step through inverting Equation~\ref{eq:basic-m-mode-analysis}. However, that demonstration
was restricted to single channel imaging due to---in part---the computational and storage
requirements associated with computing $\b B$.

\subsection{Hierarchical Transfer Matrices}

Modern interferometers are composed of large numbers of antennas ($N \gg 10$) arranged in
configurations that have both long and short baselines. For instance, the OVRO-LWA has over 30,000
baselines. The shortest baseline is 5\,m, and the longest baseline is 1.5\,km. Consequently the
OVRO-LWA measures a large range of angular scales. We can exploit this fact to reduce the computer
time and disk space required to compute and store the transfer matrix $\b B$.

The sensitivity of a baseline of length $b$ to spherical harmonic coefficients with multipole moment
$l$ is $\propto j_l(2\pi b/\lambda)$, where $j_l$ is the spherical Bessel function of the first
kind, and $\lambda$ is the wavelength. When $l \gtrsim 2\pi b/\lambda$, the spherical Bessel
functions rapidly drop to zero (see Appendix~\ref{app:spatial-to-angular} for more details about
spherical Bessel functions). Consequently, even though the transfer matrix is block-diagonal, each
diagonal block of the transfer matrix can also contain a large number of zero-elements.

Therefore, when the columns and rows of each transfer matrix block $\b B_m$ are sorted by the
multipole number $l$ and baseline length respectively, each block has the following structure:
\begin{equation}
    \b B_m = \left(
        \quad
        \begin{tikzpicture}[baseline=+13mm, scale=0.3]
            \draw [->] (0, 9) -- (7, 9);
            \node at (3.5, 10) {$l$};
            \draw [->] (8, 0) -- (8, 8);
            \node [rotate=270] at (9, 4) {baseline length};
            \fill [fill=black!70] (0, 0) rectangle (1, 1);
            \fill [fill=black!70] (0, 1) rectangle (1, 2);
            \fill [fill=black!70] (0, 2) rectangle (1, 3);
            \fill [fill=black!70] (0, 3) rectangle (2, 4);
            \fill [fill=black!70] (0, 4) rectangle (2, 5);
            \fill [fill=black!70] (0, 5) rectangle (3, 6);
            \fill [fill=black!70] (0, 6) rectangle (5, 7);
            \fill [fill=black!70] (0, 7) rectangle (7, 8);
        \end{tikzpicture}
        \quad
    \right)
\end{equation}
Shaded regions represent elements with nonzero value, whereas unshaded regions represent elements
with approximately zero value due to the fact that $l \gtrsim 2\pi b/\lambda$. This structure makes
it apparent that it is not necessary to store every element of each transfer matrix block. In fact,
by partitioning the array into sets of baselines with similar length, one can achieve significant
cost savings when computing and storing the transfer matrix elements.

Ultimately, for the OVRO-LWA we achieve a 58\% compression of the transfer matrix by not storing
elements that are approximately zero.

\subsection{Data Compression}\label{sec:compression}

Further data compression is desirable because it reduces the computational costs of all following
analysis steps. We implement the singular value decomposition (SVD) compression described by
\citep{2014ApJ...781...57S, 2015PhRvD..91h3514S}. The SVD factorizes a matrix into a unitary matrix
$\b U$, a diagonal matrix $\b\Sigma$, and another unitary matrix $\b V$ such that
\begin{equation}
    \b B = \b U \b\Sigma \b V^*\,.
\end{equation}
The diagonal elements of $\b\Sigma$ are called singular values and, in this case, represent the
amplitude of the response of the interferometer to the corresponding singular vectors (i.e., the
columns of $\b U$). The data can therefore be compressed by selecting all singular values above a
given threshold and computing
\begin{align}
    &\b R = \begin{pmatrix}
        & \vdots & \vdots & \\
        \cdots & \b u_{i} & \b u_{i+1} & \cdots \\
        & \vdots & \vdots & \\
    \end{pmatrix} \\
    &\b v_\text{compressed} = \b R^*\b v\,,
\end{align}
where $\b u_i$ is a column of $\b U$ whose singular value passes the threshold, and $\b v$ is the
vector of measured $m$-modes. The transfer matrix is similarly transformed $\b B_\text{compressed} =
\b R^*\b B$, and covariance matrices become $\b C_\text{compressed} = \b R^*\b C\b R$.

This compression is especially effective for the OVRO-LWA because the compactness of the
interferometer leads to many partial redundancies between similar baselines. This is simply a
statement that the number of baselines used in the calculation $N_\text{baselines}$ is larger than
the number of unknowns in each transfer matrix block. In this paper, we adopted $l_\text{max} = 300$
as the maximum value of the multipole number. For the OVRO-LWA $N_\text{baselines} \gg 300$, so
there are many redundancies in the dataset even though no pair of baselines is individually
redundant. In total this compression reduces the volume of data to a mere 0.6\% of its original
size (before discarding any singular values).

\section{Covariance Matrices}\label{sec:sensitivity}

We model the covariance of the observations $\b C = \langle \b v \b v^*\rangle$ with contributions
from thermal noise $\b C_\text{noise}$, foreground emission $\b C_\text{fg}$, and the cosmological
21\,cm signal itself $\b C_\text{21}$
\begin{equation}\label{eq:sum-of-covariances}
    \langle\b v\b v^*\rangle = \b C
        = \b C_\text{21}
        + \b C_\text{fg}
        + \b C_\text{noise}\,,
\end{equation}
where this expression implicitly assumes that the sky is an isotropic Gaussian-random field, and
that the sky covariance should be understood as an average over realizations of the sky.

We will begin with a detailed description of the models, measurements, and calculations used to
compute each of these covariance matrices.

\subsection{Thermal Noise Covariance}\label{sec:noise-covariance}

\begin{figure*}
    \centering
    \includegraphics[width=\textwidth]{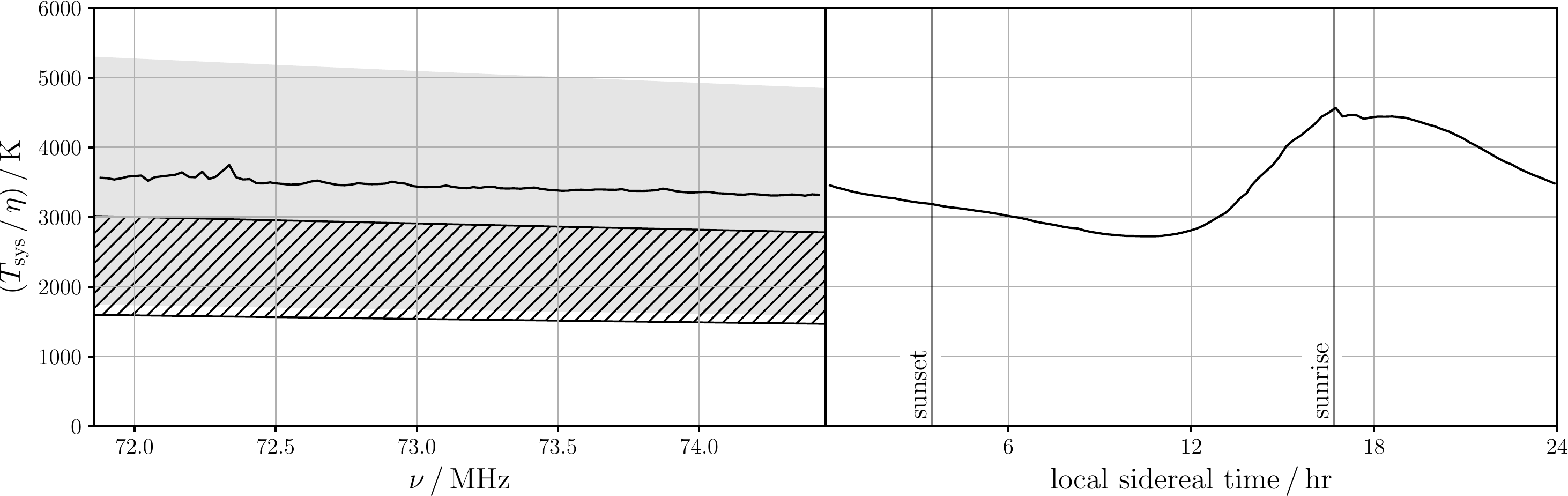}
    \caption{
        The system temperature $T_\text{sys}$ (scaled by the antenna efficiency $\eta$) measured as
        a function of frequency (left panel, solid black line), and local sidereal time (right
        panel, solid black line). The hatched region denotes the range of sky temperatures measured
        by the LEDA experiment \citep{2018MNRAS.478.4193P}. The shaded region denotes the range of
        sky temperatures measured by the EDGES experiment in the southern hemisphere
        \citep{2017MNRAS.464.4995M}.
    }
    \label{fig:Tsys}
\end{figure*}

\begin{figure*}
    \centering
    \includegraphics[width=\textwidth]{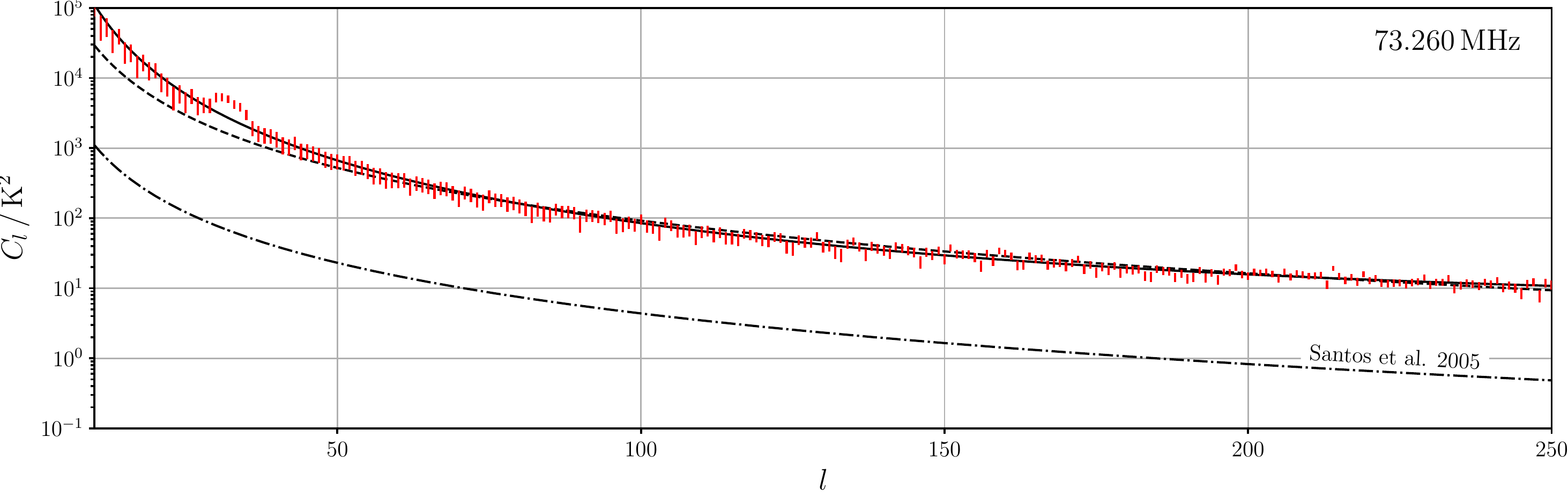}
    \caption{
        The angular power spectrum of the sky as measured by the OVRO-LWA at 73.260\,MHz.
        Measurements (with 95\% uncertainty) are indicated with red bars. The uncertainty is
        dominated by sample variance. The dashed black line is the best-fit power-law spectrum, and
        the solid black line is the best-fit solution when the power-law index is allowed to run.
        The dash-dot line is a model derived, in part, from the Haslam 408\,MHz sky map
        \citep{1981A&A...100..209H, 1982A&AS...47....1H, 2005ApJ...625..575S}.  The feature at
        $l\sim30$ is sensitive to the choice of covariance matrix, and is therefore likely
        instrumental.
    }
    \label{fig:foreground-covariance}
\end{figure*}

The 21\,cm signal is expected to be unpolarized,\footnote{
    \citet{2017PhRvD..95h3010V} find that circular polarization may be used to measure primordial
    magnetic fields, but the amplitude of this effect is too small to consider measuring with
    existing low-frequency telescopes.
} so we form
Stokes-I visibilities from the mean of the $xx$ and $yy$ visibilities. Under this convention, the
covariance of the complex-valued Stokes-I visibilities is \citep[Chapter 9]{1999ASPC..180.....T}:
\begin{equation}
    \b C_\text{noise}
        = \left(
            \frac{2 k_B T_\text{sys}}{\eta A_\text{eff} \sqrt{2\Delta\nu\tau}}
        \right)^2 \b I\,,
\end{equation}
where $k_B$ is the Boltzmann constant, $T_\text{sys}$ is the system temperature, $\eta$ is the
antenna efficiency, $A_\text{eff}$ is the effective collecting area (each assumed to be the same for
all antennas), $\Delta\nu$ is the bandwidth, and $\tau$ is the total integration time.  The
effective collecting area of the antenna is related to the solid-angle of the primary beam $\Omega$
through $A_\text{eff} = \lambda^2 / \Omega$. At 73\,MHz, the OVRO-LWA dipoles have primary beams
with $\Omega \sim 2.4\,\text{sr}$ or $A_\text{eff}\sim 7\,\text{m}^2$.

OVRO-LWA dipoles are designed to be sky-noise dominated \citep[$\ge6$\,dB between 20--80\,MHz;
][]{2012PASP..124.1090H}. More precisely, the system temperature is given by
\begin{equation}
    T_\text{sys} \approx \eta T_\text{sky} + T_\text{pre-amp}\,,
\end{equation}
where $\eta$ is the antenna efficiency, $T_\text{sky}$ is the averaged brightness temperature of the
sky (primarily the galactic synchrotron emission) weighted by the primary beam pattern, and
$T_\text{pre-amp}$ is the noise temperature of the first amplifier in the analog signal path. We
expect $T_\text{pre-amp} \approx 250\,\text{K}$ and $\eta \lesssim 0.5$ \citep{2012PASP..124.1090H}.

The LEDA experiment hosted at the OVRO-LWA measured the brightness temperature of the diffuse
galactic emission in the northern hemisphere using the five radiometric antennas
\citep{2018MNRAS.478.4193P}. At 70~MHz, the brightness temperature varies between 1700\,K and
3200\,K with a relatively flat spectral index that varies between $-2.28$ and $-2.38$.  In the
southern hemisphere, the EDGES experiment measured that the brightness temperature of the sky at
150\,MHz varies between 257\,K and 842\,K with a spectral index that varies between $-2.50$ and
$-2.62$ \citep{2017MNRAS.464.4995M}. Extrapolating to 70\,MHz, we expect the beam-weighted sky
brightness temperature in the southern hemisphere to vary between 1700\,K and 6200\,K. The maximum
brightness temperature corresponds to sidereal time when the galactic center transits.

We measured the system temperature as a function of frequency and sidereal time using a five-point
stencil to suppress the contribution of the sky emission to the measured visibilities
\begin{align*}
    \Delta(\nu,\,t) = 4V(\nu, t)
                     &- V(\nu - 24\,\text{kHz},\,t)
                      - V(\nu,\,t - 13\,\text{s}) \\
                     &- V(\nu + 24\,\text{kHz},\,t)
                      - V(\nu,\,t + 13\,\text{s})\,,
\end{align*}
where $\Delta(\nu,\,t)$ is a quantity whose variance is 20 times larger than that of the measured
visibilities $V(\nu,\,t)$ at the given frequency $\nu$ and time $t$. Note that 24\,kHz is the native
frequency resolution of the OVRO-LWA and 13\,s is the integration time.  Therefore this stencil
takes the difference between each measured visibility and the bilinear interpolation from adjacent
frequency channels and time integrations. We then estimated the system temperature from the variance
of $\Delta$. The measured system temperature is shown in Figure~\ref{fig:Tsys} compared to the sky
temperature measured by LEDA and extrapolated from EDGES.  As expected, the system temperature
increases at lower frequencies due to the increasing sky brightness temperature, and varies
sidereally reaching a maximum as the galactic center transits the meridian. These measurements
suggest that the antenna efficiency $\eta \sim 0.25$. Although the system temperature varies with
time and frequency, we adopt a constant system temperature of $3500\eta\,\text{K}$ when computing
the sensitivity of the OVRO-LWA. We expect this approximation to potentially introduce errors of
$\sim 10\%$ to the computed sensitivity and error bars, which does not materially impact the results
presented in this paper. \revision{
    In \S\ref{sec:non-stationary} we additionally simulate the impact of the implicit assumption of
    stationary thermal noise and conclude that it also does not bias our results.
}

\subsection{Foreground Covariance}\label{sec:foreground-covariance}

Under the assumption of a Gaussian random field, the covariance contributed by the sky can be
computed from the multi-frequency angular power spectrum:
\begin{equation}\label{eq:multi-frequency-angular-power-spectrum}
    \langle a_{lm}(\nu) \, a_{l^\prime m^\prime}^*(\nu^\prime)\rangle
        = C_l(\nu, \nu^\prime) \, \delta_{ll^\prime} \, \delta_{mm^\prime}\,,
\end{equation}
where the angled brackets denote an ensemble average over realizations of the sky, $a_{lm}(\nu)$ is
the spherical harmonic coefficient of the sky brightness at frequency $\nu$, $C_l(\nu, \nu^\prime)$
is the multi-frequency angular power spectrum at the multipole moment $l$, and between the
frequencies $\nu$ and $\nu^\prime$. The Kronecker delta is represented by $\delta$.  The
transfer-matrix $\b B$ describes how to relate the covariance of the spherical harmonic coefficients
to the covariance of the measurements themselves, such that
\begin{equation}\label{eq:maps-to-covariance}
    \b C_\text{sky} = \b B \b C^\prime_\text{sky} \b B^*\,,
\end{equation}
where $\b C_\text{sky}$ is a term in Equation~\ref{eq:sum-of-covariances}, and $\b
C^\prime_\text{sky}$ is a matrix whose elements are specified by
Equation~\ref{eq:multi-frequency-angular-power-spectrum}.

A common parameterization of $C_l(\nu_1, \nu_2)$ for foreground radio emission is
\citep{2005ApJ...625..575S}
\begin{eqnarray}\label{eq:cforeground}
    C_l^\text{fg}(\nu, \nu^\prime) =
    \sum_i &A_i& \left(\frac{l}{l_0}\right)^{-\alpha_i}
                 \left(\frac{\nu\nu^\prime}{\nu_0^2}\right)^{-\beta_i} \nonumber \\
           &\times&\exp\left(-\frac{(\log\nu-\log\nu^\prime)^2}{2\zeta_i^2}\right)\,,
\end{eqnarray}
where $A_i$ represents the overall amplitude of a foreground component.  $\alpha_i$ determines its
angular spectrum, and $\beta_i$ determines its frequency spectrum. Finally, $\zeta_i$ controls the
degree to which nearby frequency channels are correlated. The statement that foreground emission is
spectrally smooth here implies \revision{$\zeta_i^2 \gg \log^2(\nu/\nu^\prime)$ for each component.
This parameterization allows for multiple power-law foreground components and ensures that the
covariance matrix is positive definite.  Because the fractional bandwidth is small, in this paper we
assume $\zeta_i^2 \gg \log^2(\nu/\nu^\prime)$. For simplicity when measuring the foreground
covaraiance, we will additionally assume a single foreground component such that
Equation~\ref{eq:cforeground} can be written as}
\begin{equation}
    C_l^\text{fg}(\nu, \nu^\prime) = \sqrt{C_l^\text{fg}(\nu)\,C_l^\text{fg}(\nu^\prime)}\,,
\end{equation}
where $C_l^\text{fg}(\nu) = C_l^\text{fg}(\nu, \nu)$ is the single-frequency angular power spectrum.

We measured the angular power spectrum of the foreground emission at each frequency channel using a
quadratic estimator \citep{1997PhRvD..55.5895T}. The angular power spectrum is given by
\begin{equation}
    C_l^\text{fg}(\nu) = \left[\b F^{-1} (\b q - \b b)\right]_l\,,
\end{equation}
where $\b F$ is the Fisher information matrix, $\b q$ is a quadratic function of the input data, and
$\b b$ is the bias due to thermal noise.  \revision{The elements of the Fisher matrix $\b F$ are
given by
\begin{equation}
    F_{ll^\prime} = \sum_m \left| \b w_{lm}^*\b C_m^{-1} \b w_{l^\prime m} \right|^2\,,
\end{equation}
where $F_{ll^\prime}$ is the Fisher matrix element corresponding to the multipole numbers $l$ and
$l^\prime$, $\b w_{lm}$ is the column of the transfer matrix corresponding to $l$ and the azimuthal
quantum number $m$, and $\b C_m$ is the covariance matrix block corresponding to $m$. $|\cdot|$ is
used here to indicate the magnitude of a complex number. The elements of $\b q$ and $\b b$ are given
by
\begin{align}
    q_l &= \sum_m \left| \b w_{lm}^*\b C_m^{-1} \b v_m \right|^2 \\
    b_l &= \sum_m \left\| \b w_{lm}^*\b C_m^{-1} \b C^{1/2}_{\text{noise}, m} \right\|^2\,,
\end{align}
where $\b v_m$ is the vector of $m$-modes corresponding to the given value of $m$, and $\b
C_{\text{noise}, m}$ is the corresponding block of the noise covariance matrix. $\|\cdot\|$ is
used here to indicate the magnitude of a complex vector (the usual Euclidean norm).}

The result of applying this quadratic estimator to the dataset at 73.260\,MHz (a representative
channel) can be seen in Figure~\ref{fig:foreground-covariance}. Broadly, the data can be described
with a power law in $l$, but the quality of the fit is somewhat poor. A single power-law fit gives
\begin{equation}
    C_l \sim 92. \times \left(\frac{l}{100}\right)^{-2.5} \,\text{K}^2\,.
\end{equation}
In fact, while this is a reasonable fit at $l > 75$, a shallower power-law index is preferred $l <
75$. If we allow for the power-law index to run, the best-fit model becomes:
\begin{equation}\label{eq:measured-cforeground}
    C_l \sim 85. \times \left(\frac{l}{100}\right)^{-3.2 + l/277.} \,\text{K}^2\,.
\end{equation}
A comparison of these two models can be seen in Figure~\ref{fig:foreground-covariance} in addition
to a model of the galactic synchrotron emission derived by \citet{2005ApJ...625..575S}, which
appears to underestimate the amplitude of $C_l$ by an order of magnitude.  Because the fractional
bandwidth of this measurement is small, essentially all reasonable spectral indices are permitted.
We adopt a fiducial spectral index of $-2.5$ as a compromise between the spectral indices measured
by LEDA and EDGES. \revision{We will discuss the potential impact of an error in the spectral index
in \S\ref{sec:calibration-errors} as part of a broader discussion of calibration and bandpass
errors.}

\subsection{Signal Covariance}\label{sec:signal-covariance}

\begin{figure}[t]
    \centering
    \includegraphics[width=\columnwidth]{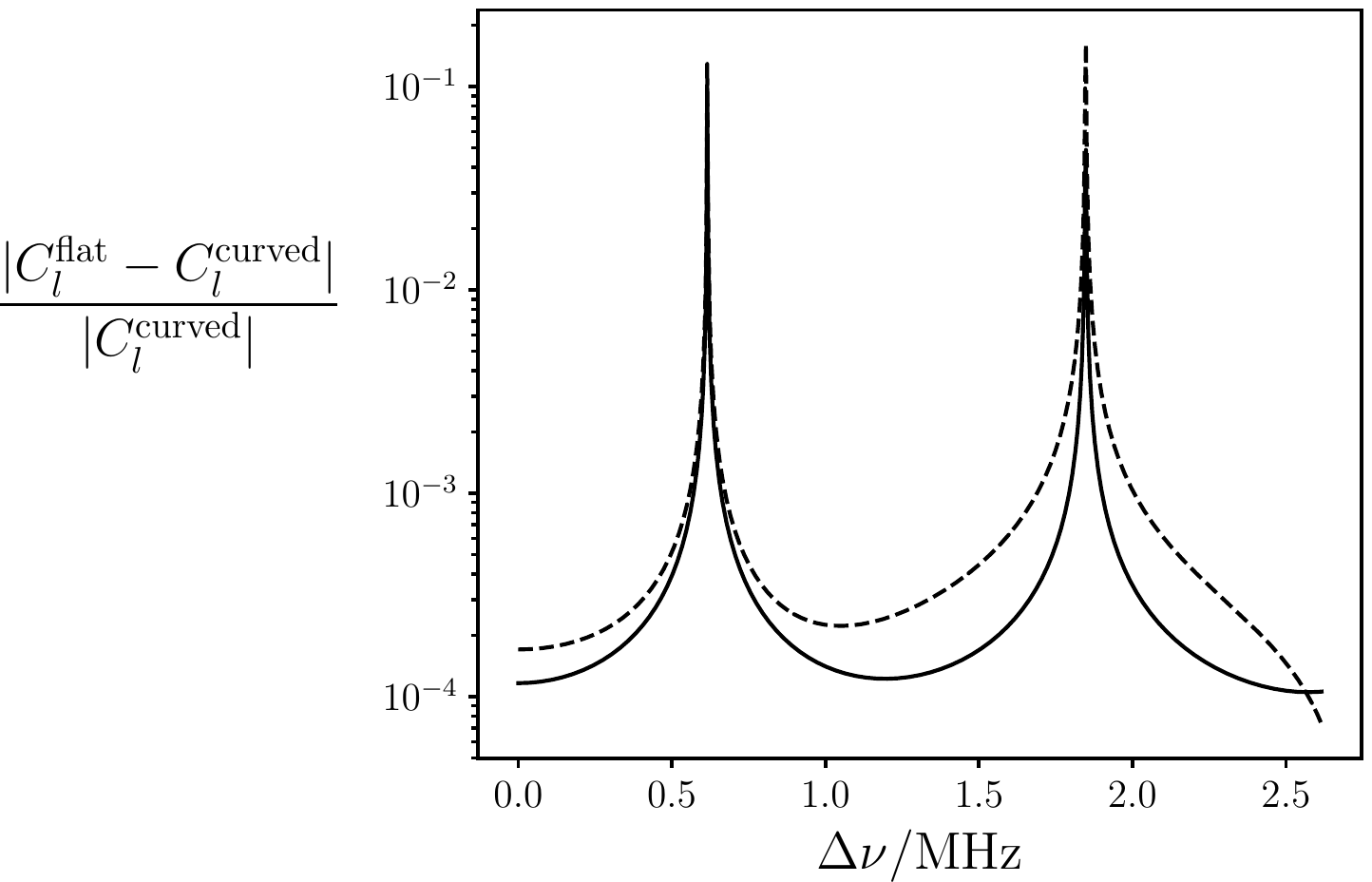}
    \caption{
        The relative error involved with making the flat-sky approximation for a hat function power
        spectrum (i.e., the relative difference between Equations~\ref{eq:csignal-curved-sky} and
        \ref{eq:csignal-flat-sky}) with $l=10$ (solid line) and $l=100$ (dashed line). The hat
        function is centered at $k_\parallel=0.1\,\text{Mpc}^{-1}$ with a domain that extends from
        $0.095\,\text{Mpc}^{-1}$ to $0.105\,\text{Mpc}^{-1}$. The spikes in relative error
        correspond to when $C_l^\text{curved}(\Delta\nu) \approx 0$.
    }
    \label{fig:flat-sky-approximation}
\end{figure}

Given the isotropic three-dimensional spatial power spectrum of the 21\,cm brightness temperature
$P^{21}_z(k)$ with the wavenumber $k$ and at the redshift $z$, the multi-frequency angular
power spectrum $C_l(\nu, \nu^\prime)$ is given by
\begin{equation}\label{eq:csignal-curved-sky}
    C^{21}_l(\nu, \nu^\prime) =
        \frac{2}{\pi}
        \int
        P_z^{21}(k) \,
        j_l(k r_z) \,
        j_l(k r_{z^\prime}) \,
        k^2 \, \d k\,,
\end{equation}
where $r_z$ is the comoving distance to the redshift $z$ (specified by the frequency $\nu$), and
$j_l(x)$ is the spherical Bessel function of the first kind. In the flat-sky approximation,
Equation~\ref{eq:csignal-curved-sky} can be simplified to
\begin{equation}\label{eq:csignal-flat-sky}
    C^{21}_l(\nu, \nu^\prime) \approx
        \frac{1}{\pi r_z r_{z^\prime}}
        \int
        P_z^{21}(k_\perp, k_\parallel) \,
        \cos\left(k_\parallel \Delta r_z\right)
        \, \d k_\parallel\,,
\end{equation}
where $k_\perp = l/r_z$ and $k_\parallel = \sqrt{k^2-k_\perp^2}$.  See
Appendix~\ref{app:spatial-to-angular} for a derivation of this approximation and the assumptions
that must be satisfied for it to be a reasonable approximation.

If $P_z^{21}(k_\perp, k_\parallel)$ is additionally assumed to be a piece-wise linear function,
Equation~\ref{eq:csignal-flat-sky} can be evaluated analytically. Under this assumption,
$P_z^{21}(k_\perp, k_\parallel)$ can be represented using linear hat functions (triangular functions
in two dimensions), such that
\begin{align}
    P_z^{21}(k_\perp, k_\parallel) &= \sum_\alpha p_\alpha
        \times {\rm hat}_\alpha(k_\perp, k_\parallel)
        \label{eq:palpha} \\
    C^{21}_l(\nu, \nu^\prime) &\approx
        \frac{1}{\pi r_z r_{z^\prime}}
        \sum_\alpha p_\alpha H_\alpha(\Delta r_z)
\end{align}
where $H_\alpha(\Delta r_z) = \int {\rm hat}_\alpha(k_\perp, k_\parallel) \, \cos\left(k_\parallel
\Delta r_z\right) \, \d k_\parallel$.

The flat-sky approximation is valid only when the power spectrum is smooth enough for rapid
oscillations in the spherical Bessel functions to cancel out. The hat functions are
non-differentiable, and so we must compute the error associated with this pixelization of the power
spectrum. Figure~\ref{fig:flat-sky-approximation} gives the relative error on the computed angular
power spectrum for a fiducial hat function power spectrum. Generally the error is $10^{-4}$, but can
reach to $10^{-1}$ at values where $C_l \approx 0$. This is an acceptable error in the context of
this paper, but future experiments may wish to experiment with differentiable basis functions.

When selecting a fiducial model for the 21\,cm power spectrum we prefer to remain unopinionated, and
therefore adopt a flat power spectrum with a single free parameter, the overall amplitude of the
dimensionless power spectrum $\Delta_{21}$:
\begin{equation}
    P_\text{fiducial}^{21}(k) = \frac{2\pi^2}{k^3}\Delta_{21}^2\,.
\end{equation}
Prior to the recent detection of an absorption feature centered at 78\,MHz by
\citet{2018Natur.555...67B}, the amplitude of the power spectrum was generally predicted to be
$\Delta_{21} < 20\,\text{mK}$ at $z\sim 20$ \citep[e.g.,][]{2014MNRAS.437L..36F}. However, more
recent predictions in the context of the measured 78\,MHz absorption feature predict a much brighter
power spectrum \citep[e.g.][]{2018Natur.555...71B, 2018arXiv180503254K}. However, we adopt
$\Delta_{21} = 20\,\text{mK}$ as a fiducial power spectrum amplitude in order to be somewhat
conservative.

The amplitude of the fiducial 21\,cm signal is primarily used to determine which modes should be
kept by the foreground filter described in the following section. Therefore, if the reader is
skeptical of this selection of the power spectrum amplitude, they may simply choose to interpret the
results as if the foreground filter is stronger or weaker than expected. A foreground filter that is
weaker than expected could lead to a biased power spectrum estimate, but a foreground filter that is
stronger than expected will lead to a conservative result with error bars that may be larger than
necessary.

\section{Foreground Filtering}\label{sec:foreground-filtering}

\begin{figure*}
    \centering
    \includegraphics[width=\textwidth]{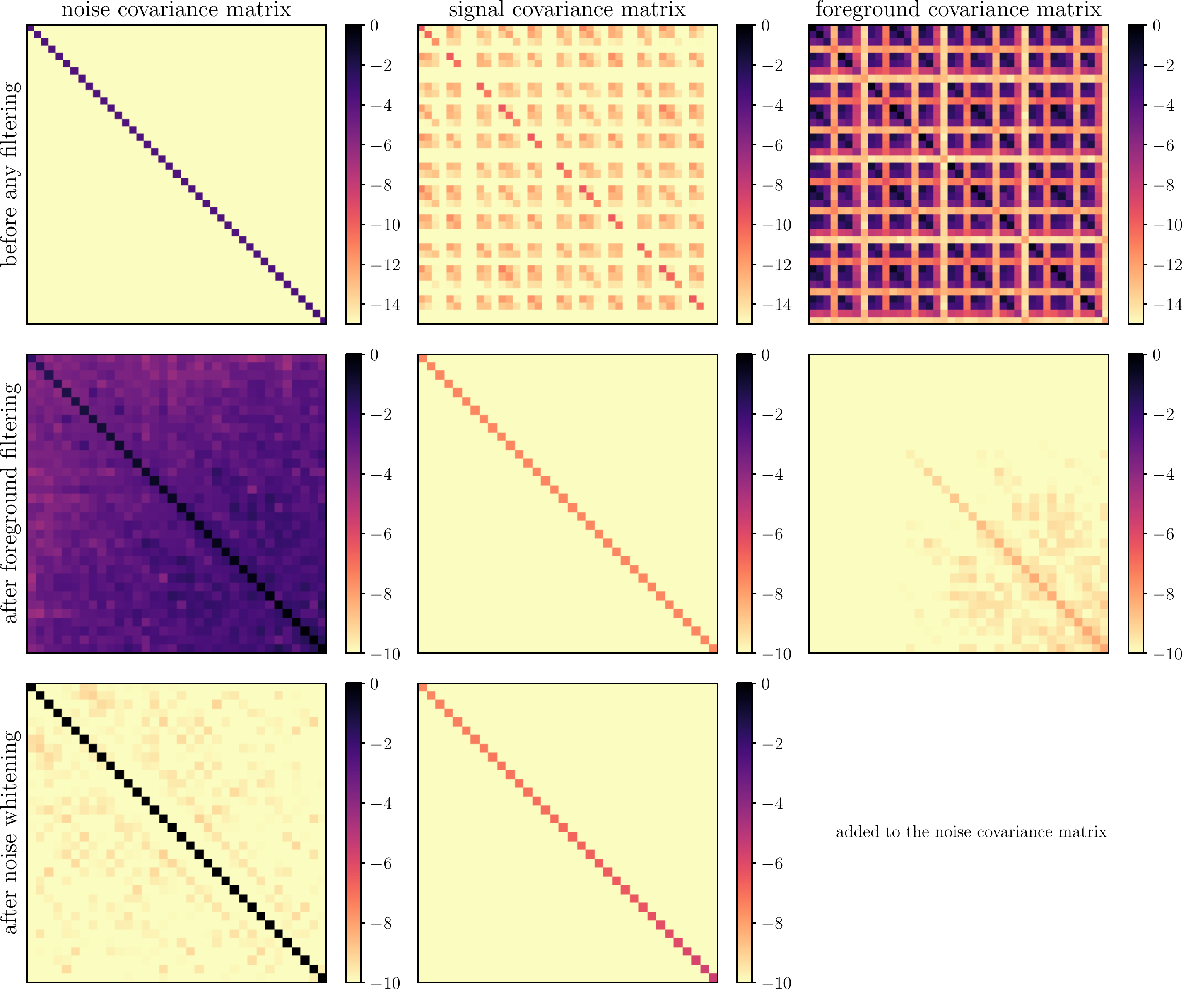}
    \caption{
        Illustration of the action of foreground filtering on each of the covariance matrices
        discussed in \S\ref{sec:sensitivity}. The left column corresponds to the noise covariance
        matrix, the middle column corresponds to the high-redshift 21\,cm contribution to the
        covariance, and the right column corresponds to the foreground covariance matrix. The top
        row is before any filtering has been applied, the middle row is after the first KL
        transform, and the bottom row is after the second KL transform. \revision{The color gives
        the logarithm (base 10) of the absolute value of the matrix element. Prior to any
        filtering, baselines are ordered by length and then by frequency. The apparent structure in
        the signal and foreground covariances is due to this ordering combined with the downsampling of the
        matrices necessary to produce this graphic. After filtering and whitening, the rows/columns
        are ordered by the magnitude of the corresponding eigenvalues. This diagram illustrates which
        matrices are diagonal, and the relative amplitude of the matrix elements after each stage in
        the processing.}
    }
    \label{fig:foreground-filtering-illustration}
\end{figure*}

\begin{figure}[t]
    \centering
    \includegraphics[width=\columnwidth]{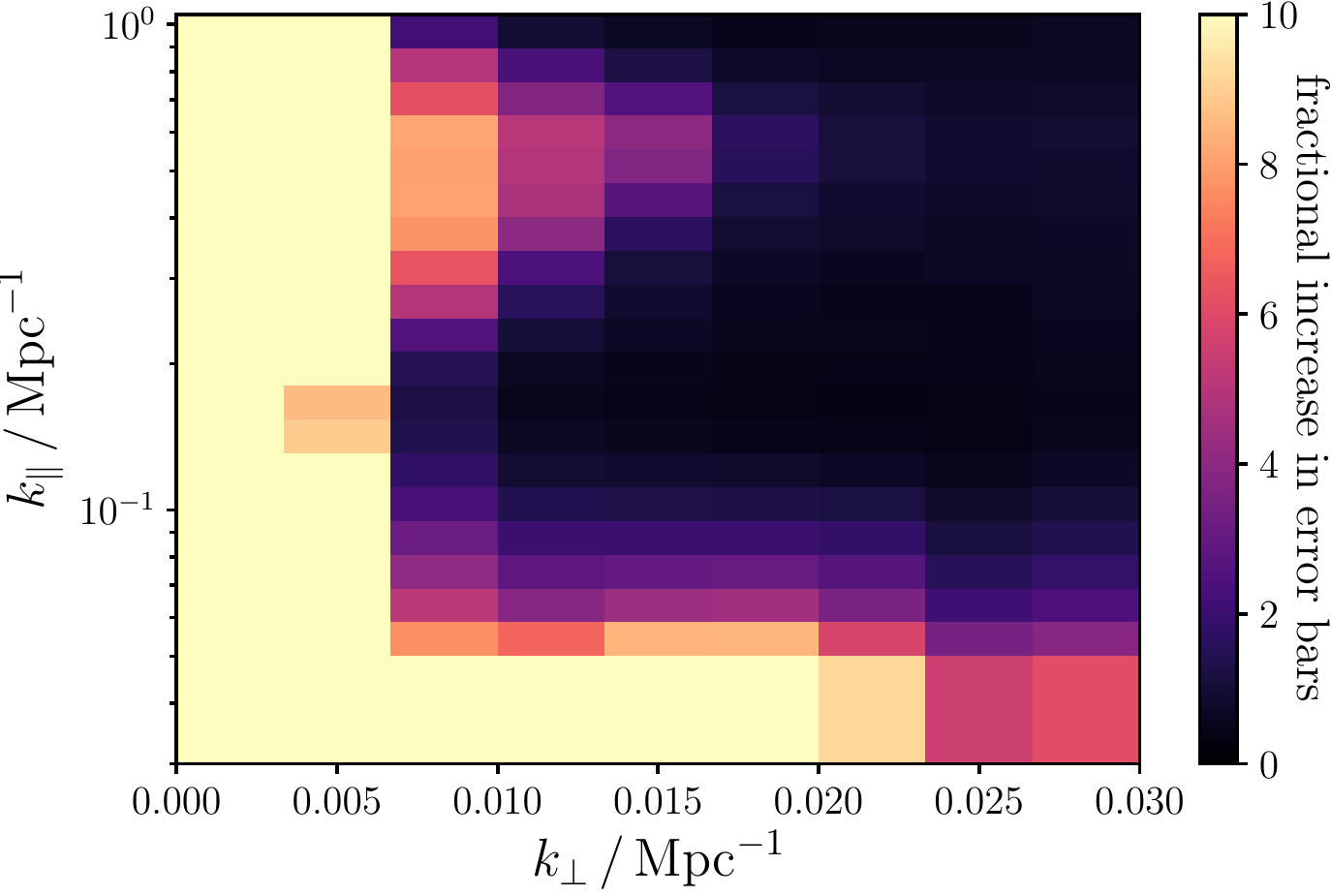}
    \caption{
        The fractional increase in the size of the error bars in each power spectrum bin due to the
        application of a double KL transform foreground filter (moderate strength).
    }
    \label{fig:error-bars-increase}
\end{figure}

\begin{figure}[t]
    \centering
    \includegraphics[width=\columnwidth]{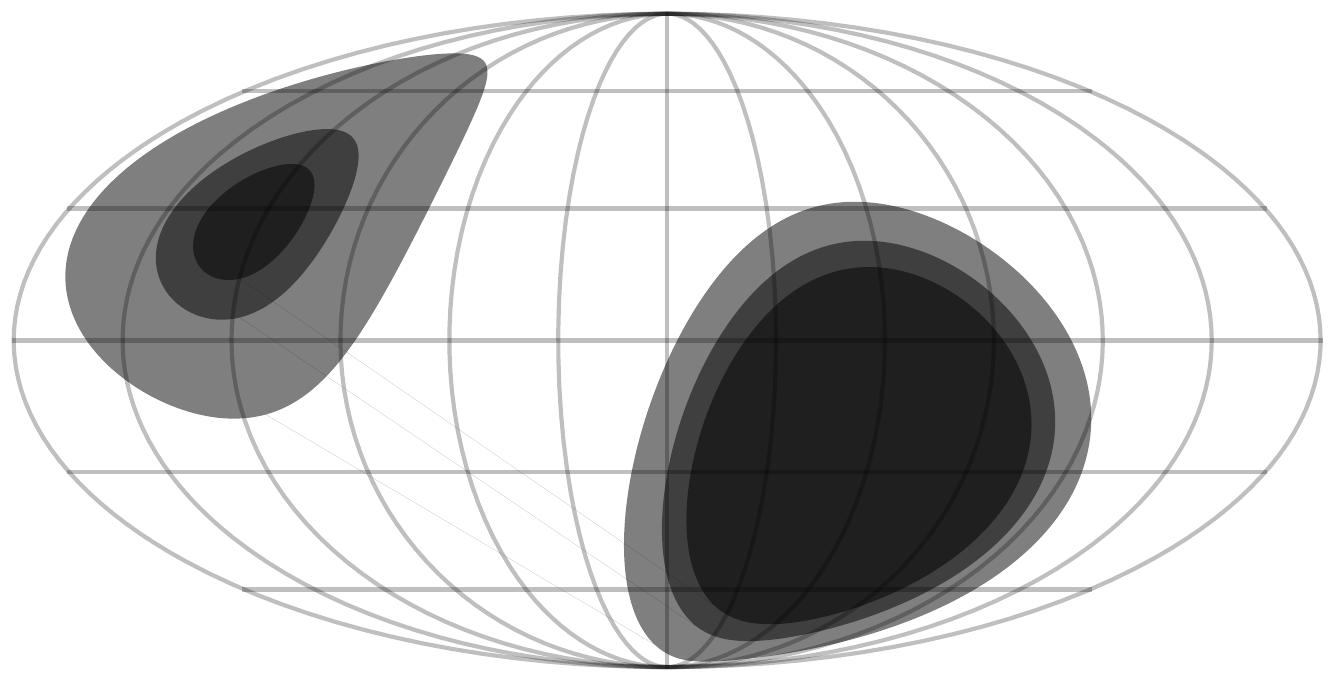}
    \caption{
        Mollweide projected illustration of the sky where shaded regions are down-weighted by the
        foreground filter. \revision{More precisely, we measured the RMS along rings of constant
        declination after foreground filtering. Shaded regions are where the RMS is less than 1\% of
        the maximum RMS.} From darkest to lightest, these regions of the sky are filtered by the mild,
        moderate, and extreme foreground filters respectively.
    }
    \label{fig:foreground-filtering-schematic}
\end{figure}

In the preceding sections, we have derived and---where appropriate---measured the contribution of
thermal noise, foreground emission and the cosmological 21\,cm emission to the complete covariance
matrix of the data. This was possible because the transfer matrix $\b B$ is block-diagonal with
respect to $m$, and we assumed that the sky emission is a Gaussian-random field (i.e., there are no
correlations between different values of $m$). Without these properties the full covariance matrix
is generally too large to represent and manipulate on any existing computer.
\citet{2014ApJ...781...57S, 2015PhRvD..91h3514S} were therefore able to derive a new foreground
filtering technique that exploits knowledge of the full covariance matrix. This filter is called the
double Karhunen--Lo\`{e}ve transform (double KL transform). In this section we will briefly
summarize the action of this foreground filter and demonstrate its application to the OVRO-LWA. We
will finally attempt to develop an intuitive understanding by relating its behavior to the
``foreground wedge'' commonly seen in the literature
\citep[e.g.,][]{2012ApJ...745..176V,2012ApJ...756..165P,2015ApJ...804...14T}.

The KL transform is closely related to the generalized eigenvalue problem. For two Hermitian,
positive definite matrices $\b C_1, \b C_2 \in \mathbb{C}^N$, we would like to find all pairs of
eigenvalues
$\lambda_i$ and eigenvectors $\b v_i$ for which
\begin{equation}
    \b C_1 \b v_i = \lambda_i \b C_2 \b v_i\,.
\end{equation}
Because both matrices are Hermitian, it quickly follows that the eigenvalues $\lambda_i$ must be
real. Because both matrices are additionally positive definite, it follows that the eigenvalues
$\lambda_i$ must all be positive. Furthermore we can select the normalization of the eigenvectors
such that
\begin{align}
    \b v_i^* \b C_1 \b v_i &= \lambda_i \\
    \b v_i^* \b C_2 \b v_i &= 1\,.
\end{align}
Under this convention the eigenvalues have a simple interpretation as the ratio of the mode-power
contained in $\b C_1$ relative to $\b C_2$. All $N$ eigenvalues and eigenvectors can be conveniently
found with a single call to \textsc{LAPACK} \citep{Anderson:1990:LPL:110382.110385}.

In \S\ref{sec:foreground-covariance} we derived and measured a model for the foreground contribution
to the data covariance $\b C_\text{fg}$. In \S\ref{sec:signal-covariance} we projected a fiducial
model 21\,cm power spectrum to a multi-frequency angular power spectrum, and therefore derived its
contribution to the data covariance $\b C_\text{21}$. We can solve the generalized eigenvalue
problem for the eigenvectors (arranged as columns within the matrix $\b L$) that simultaneously
diagonalize both matrices (called the KL transform):
\begin{align}
    \b L\b C_\text{fg}\b L^* &= \b\Lambda \\
    \b L\b C_\text{21}\b L^* &= \b I \,,
\end{align}
where $\b\Lambda$ is a diagonal matrix, and $\b I$ is the identity matrix. The foreground filter is
simply constructed by selecting only the eigenvectors for which the corresponding eigenvalue (i.e.,
the foreground--signal power ratio) is less than some value $\epsilon_\text{filter}$ selected by the
observer. The application of this filter to a fiducial set of models can be seen in the second row
of Figure~\ref{fig:foreground-filtering-illustration}. The signal covariance matrix has been
diagonalized and the power in each remaining mode is greater than the surviving power in the
foreground covariance matrix. The off-diagonal elements in the foreground covariance matrix are due
to numerical errors. The possible effect of these numerical errors on the efficacy of the foreground
filter is noted here, but is out of the scope of the current work.

Much emphasis has been placed on maintaining the integrity of the ``foreground wedge'' in the next
generation of 21\,cm telescopes. In its simplest form, the existence of the foreground wedge is a
statement that most foreground radio emission that observers have to contend with when trying to
detect the cosmological 21\,cm is spectrally smooth. A simple Fourier transform of an image cube
therefore leads to most contamination occupying the space where $k_\parallel$ (the line of sight
wavenumber) is small. However, due to the chromatic nature of interferometers (specifically that the
fringe spacing $\propto b/\lambda$ where $b$ is the baseline length and $\lambda$ is the
wavelength), this contamination is spread out into a wedge-like structure. Additional chromaticity
in, for example, the bandpass or antenna primary beam leads to the contamination even leaking out of
the wedge. In the event of too much leakage, the observer has lost their ability to measure the
cosmological 21\,cm transition.

In contrast, the KL transform automatically finds the optimal linear combination of the dataset for
separating foregrounds using all available information built into the models. This includes
information on the frequency spectrum of the foregrounds as well as their angular structure, which
can lead to scenarios where the KL transform can filter foreground emission that cannot be avoided
with a delay filter. There is, of course, a caveat that the KL transform requires sufficiently
detailed models for the instrument and foreground emission. However, it is not necessarily optimal
to remain completely apathetic to the structure of foreground emission, and most collaborations are
expending significant effort to characterize their instruments.

A single KL transform, however, leads to large off-diagonal elements in the noise covariance matrix
(see the second row of Figure~\ref{fig:foreground-filtering-illustration}). Therefore
\citet{2014ApJ...781...57S,2015PhRvD..91h3514S} introduced a second KL transform that diagonalizes
the noise covariance matrix. This second matrix composed of eigenvectors will be denoted with $\b
W$. In total we therefore have
\begin{equation}
    \b C_\text{filtered}
        = \underbrace{\b W^*\b L^*\b C_\text{21}\b L\b W}_{\b S}
        + \underbrace{\b W^*\b L^*(\b C_\text{fg} + \b C_\text{noise})\b L\b W}_{\b I}\,,
\end{equation}
where $\b C_\text{filtered}$ is the data covariance matrix after applying the double KL transform
foreground filter, $\b S$ is a real diagonal matrix, and $\b I$ is the identity matrix. The diagonal
elements of $\b S$ give the expected signal--noise ratio in each mode.  The foreground filter is
applied to the measured $m$-modes by simply computing
\begin{equation}
    \b v_\text{filtered} = \b W^*\b L^*\b v\,.
\end{equation}

In this paper we will repeat the analysis using three different values for the foreground filtering
\revision{foreground--signal} threshold $\epsilon_\text{filter}$. This will allow us to assess the performance
of the foreground filter and degree to which residual foreground contamination may be affecting the
measurement. We will adopt the terminology ``strong,'' ``moderate,'' and ``mild'' to mean:
\begin{align*}
    \epsilon_\text{filter} &= 0.1 & \text{(``strong'' foreground filtering)} \\
    \epsilon_\text{filter} &= 1   & \text{(``moderate'' foreground filtering)} \\
    \epsilon_\text{filter} &= 10  & \text{(``mild'' foreground filtering)}\,.
\end{align*}
Now we will build a physical intuition for understanding the operation of the double KL transform
foreground filter.

Figure~\ref{fig:error-bars-increase} illustrates the fractional increase in error bars associated
with applying the moderate foreground filter. In the space of a cylindrically binned power spectrum,
the action of the filter is to discard linear combinations of the dataset with low $k_\parallel$ and
low $k_\perp$. This manifests itself as a decrease in sensitivity---equivalently an increase in the
error bars---in this region of parameter space.  High $k_\parallel$ modes are computed from rapid
frequency differences, whereas low $k_\parallel$ modes are slowly varying in frequency. Because the
foreground emission is spectrally smooth, it tends to corrupt modes with low $k_\parallel$. The
pattern of this contamination is known as the foreground wedge. However, the foreground filter
additionally removes emission on large angular scales (low $k_\perp$). This arises because the
foreground filter is aware that the foreground emission is brighter on larger angular scales (see
Figure~\ref{fig:foreground-covariance} and Equation~\ref{eq:measured-cforeground}).

As illustrated in Figure~\ref{fig:foreground-filtering-schematic}, the foreground filter also tends
to remove emission in two separate parts of the sky: low declinations that are never seen at high
elevations from the OVRO-LWA, and high declinations around the North Celestial Pole (NCP).  This
filtering of high and low declinations can be seen in
Figure~\ref{fig:spherical-power-spectra-filter-strength}, which is a Tikhonov-regularized image of
the sky constructed from the post-filtered data.

The OVRO-LWA is a zenith pointing drift-scanning instrument.  Therefore foreground emission located
far from zenith has a large path difference between antennas. This large path difference leads to
additional frequency structure that allows the foreground emission to contaminate higher values of
$k_\parallel$ \citep{2012ApJ...752..137M}.  Similarly, \citet{2015ApJ...804...14T} derived the
impact of widefield effects on the foreground contamination and found that baseline foreshortening
can lead to additional galactic synchrotron emission on large angular scales contaminating the
measurement. This foreground emission from low-elevations is problematic. The double KL transform
suppresses the contribution of these low elevations to the measurement.

Emission from the vicinity of the NCP is characterized by its low fringe-rate. As the Earth rotates,
emission located here moves slowly through the fringes of the interferometer. Therefore this
emission is predominantly characterized by low values of $m$. The foreground emission, however, is
brightest relative to the cosmological 21\,cm emission at low values of $l$ (large angular scales).
Because $m \le l$ for a given value of $l$, low values of $m$ are disproportionately contaminated by
the brightest diffuse components of the foreground emission. In fact, for the fiducial foreground
and signal models presented in \S\ref{sec:foreground-covariance} and \S\ref{sec:signal-covariance}
respectively, the the foreground--signal ratio of the most favorable mode is $\propto m^{-3.5}$.
This is a reflection of the fact that emission with a higher fringe rate tends to be smaller in
angular extent.  Consequently, the foreground filter aggressively discards information from small
values of $m$ and the emission located at the NCP is collateral damage because it can be difficult
to separate from the diffuse foreground emission. This can be seen in
Figure~\ref{fig:foreground-filtering-schematic} where increasing the strength of the foreground
filter increases the area around the NCP that is down-weighted.

\section{Results and Error Analysis}\label{sec:results}

\begin{figure*}
    \begin{tabular}{c}
        \includegraphics[width=\textwidth]{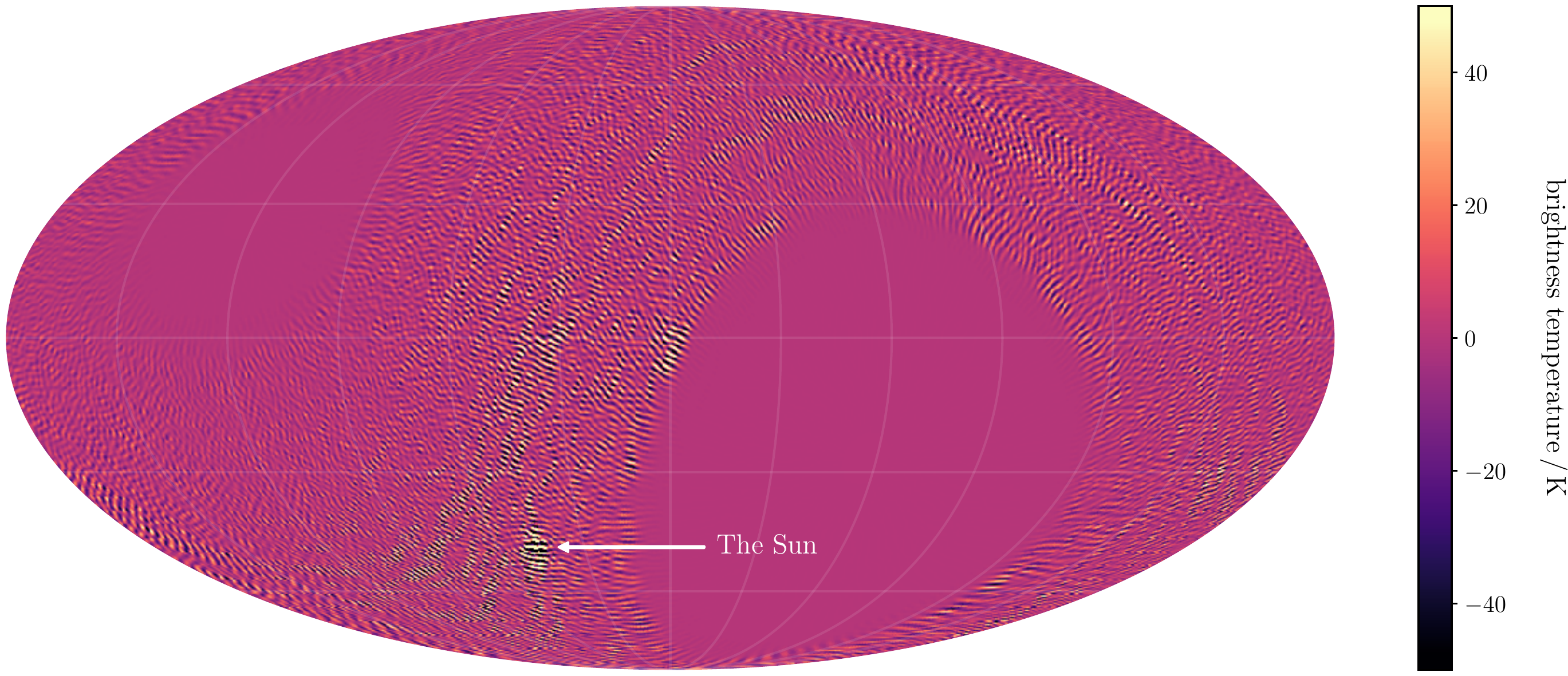}\\
        \includegraphics[width=\textwidth]{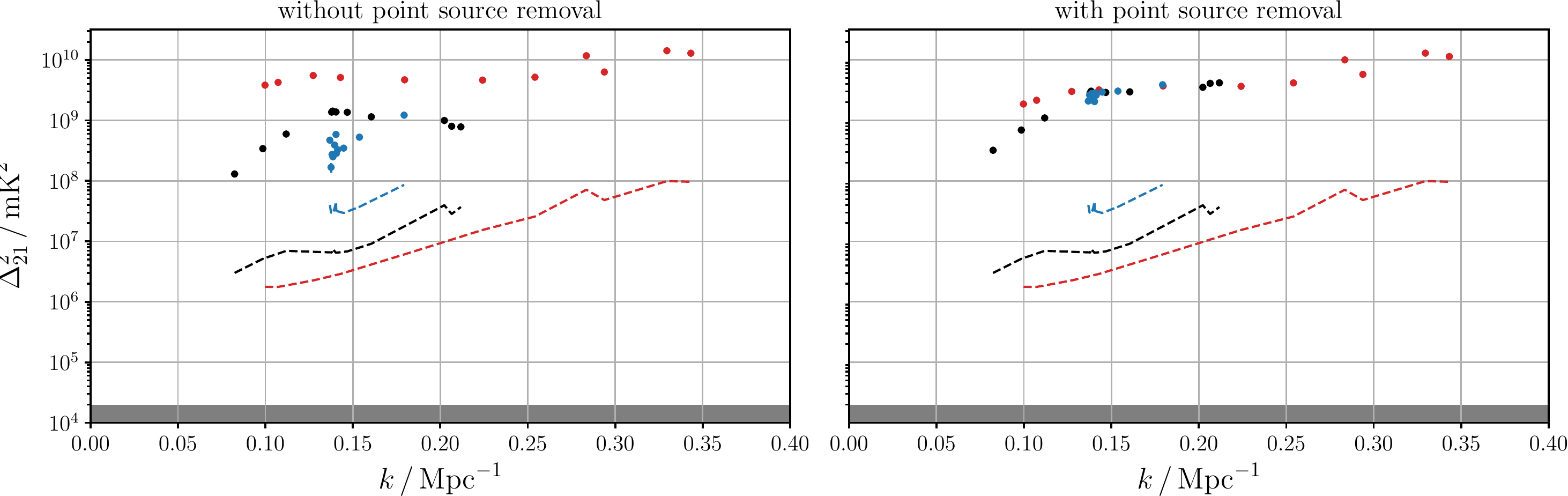}\\
    \end{tabular}
    \caption{
        (top) Mollweide projected image of the sky after point source removal and moderate
        foreground filtering. The dominant residual feature in the residuals is associated with the
        Sun.
        (bottom) The power spectrum estimated without point source removal (left) and with point
        source removal (right) at a range of filter strengths. Points correspond to the estimated
        power spectrum amplitude and the dashed lines correspond to the computed thermal noise (95\%
        confidence).  Mild foreground filtering is red, moderate foreground filtering is black, and
        extreme foreground filtering is blue.
        \revision{The shaded region represents the sensitivity
        required to detect or rule out optimistic models for the 21\,cm power spectrum.}
    }
    \label{fig:spherical-power-spectra-filter-strength}
\end{figure*}

We will use a quadratic estimator to measure the spatial power spectrum of 21\,cm fluctuations
\citep{1997PhRvD..55.5895T}. In particular we estimate the coefficients $p_\alpha$, which are
defined in Equation~\ref{eq:palpha}. As described by \citet{2003NewA....8..581P}, the observer may
tune the estimator by selecting a windowing function that produces desired properties. For example,
given the measured data $\b v$, the full covariance matrix $\b C$, and the Fisher information matrix
$\b F$, the unwindowed and minimum variance estimates of the power spectrum amplitude are
\begin{align}
    \hat{p}_\alpha^\text{unwindowed} &= \sum_\beta [\b F^{-1}]_{\alpha\beta}\,(q_\beta-b_\beta) \\
    \hat{p}_\alpha^\text{min. variance} &= \left(\sum_\beta
        F_{\alpha\beta}\right)^{-1}(q_\beta-b_\beta)\,,
\end{align}
where
\begin{align}
    q_\alpha &= \b v^*\b C^{-1}\frac{\partial\b C}{\partial p_\alpha}\b C^{-1}\b v \\
    b_\alpha &= \tr\left(\b C^{-1}\frac{\partial\b C}{\partial p_\alpha}
                         \b C^{-1}\b C_\text{noise}\right) \\
    F_{\alpha\beta} &= \tr\left(\b C^{-1}\frac{\partial\b C}{\partial p_\alpha}
                                \b C^{-1}\frac{\partial\b C}{\partial p_\beta}\right)\,.
\end{align}
Directly computing $F_{\alpha\beta}$ from its definition is computationally expensive, and so we
compute an approximation of the Fisher information matrix using the iterative Monte Carlo scheme
described by \citet{2003NewA....8..581P, 2013PhRvD..87d3005D}.

We will make exclusive use of the minimum variance estimator in this paper because it is relatively
insensitive to errors in the Fisher information matrix, which are inevitable due to the Monte Carlo
computation. Additionally, the unwindowed estimator can compound numerical errors when the condition
number of $\b F$ is large.\footnote{
    The condition number of a matrix $\b A$ is $\kappa(\b A) = \|\b A\|\,\|\b A^{-1}\|$ and
    describes the error introduced when solving the linear equation $\b A\b x=\b b$ for the vector
    $\b x$. As a general rule of thumb, if $\log_{10}\kappa(\b A) = N$, one can expect to lose $N$
    digits of precision after computing $\b A^{-1}\b b$.
}

In Figure~\ref{fig:spherical-power-spectra-filter-strength} we present the results of the quadratic
estimator with and without point source removal, and across the range of foreground filter
strengths. These estimates are, across the board, severely limited by systematic errors. This is
readily apparent due to the extreme amplitude of the estimated power.  We therefore interpret these
measurements as upper limits $\Delta_{21}^2 \lesssim (10^4\,\text{mK})^2$ at $k\approx
0.10\,\text{Mpc}^{-1}$.

As the strength of the foreground filter is increased more information is lost by the filter. This
is seen in the window functions of the quadratic estimator. With mild foreground filtering, the
window functions are roughly evenly spaced between $k=0.10\,\text{Mpc}^{-1}$ and
$0.35\,\text{Mpc}^{-1}$. With extreme foreground filtering, all of the measurements are instead
concentrated around $k=0.15\,\text{Mpc}^{-1}$, which reflects the loss of information at other
values of the wavenumber $k$.

After initial calibration and stationary component removal, we attempted to subtract the eight
brightest point sources in the northern hemisphere in addition to the Sun. The brightest of these
sources were removed with direction-dependent calibrations. The fainter sources were simply
subtracted after fitting for their flux and position (attempting to account for ionospheric
scintillation and refraction). The Sun was removed using a resolved source model. With mild
foreground filtering, this point source removal leads to a $\sim2\times$ reduction in the power
spectrum amplitude.

However, the efficacy of the foreground filter materially differs between the datasets where bright
point sources have and have not been removed. Without point source removal, increasing the strength
of the foreground filter leads to a reduction of the estimated power. This reflects the fact that
the foreground filter is removing increasing amounts of foreground contamination. In contrast, if
point sources have been subtracted, the power spectrum amplitude is insensitive to the strength of
the foreground filter. While the point source removal routine leads to less foreground contamination
in the absence of foreground filtering, it also restricts the effectiveness of the foreground
filter. This suggests that the point source removal routine introduces additional errors into the
dataset that inhibit the action of the foreground filter.

We will now attempt to diagnose the source of these residual systematic errors that limit this
measurement. While doing this, we will adopt the moderate foreground filter as the fiducial
foreground filter due to its action as a compromise between the amount of foreground emission
removed and resolution in the wavenumber $k$.

\subsection{Even--Odd Jackknife}

\begin{figure*}
    \centering
    \begin{tabular}{c}
        \includegraphics[width=\textwidth]{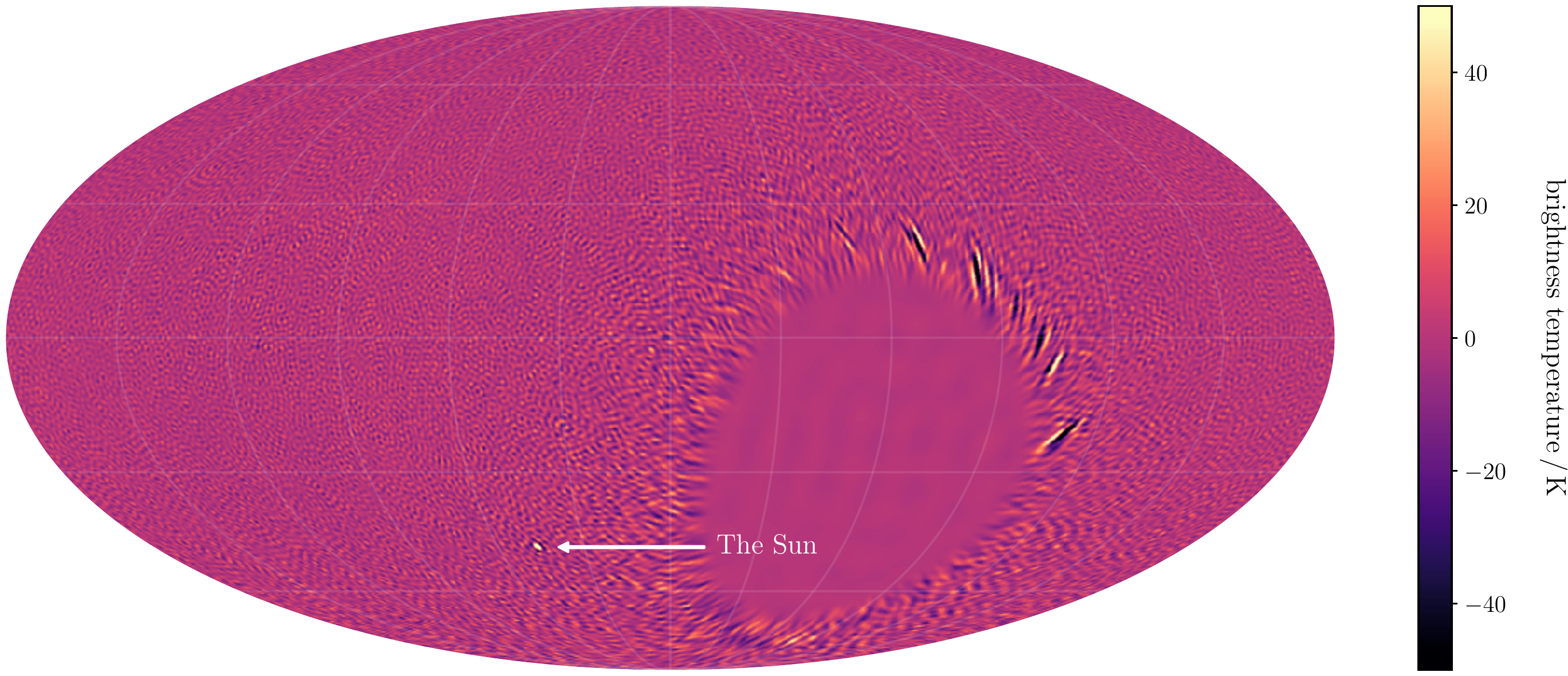}\\
        \includegraphics[width=\textwidth]{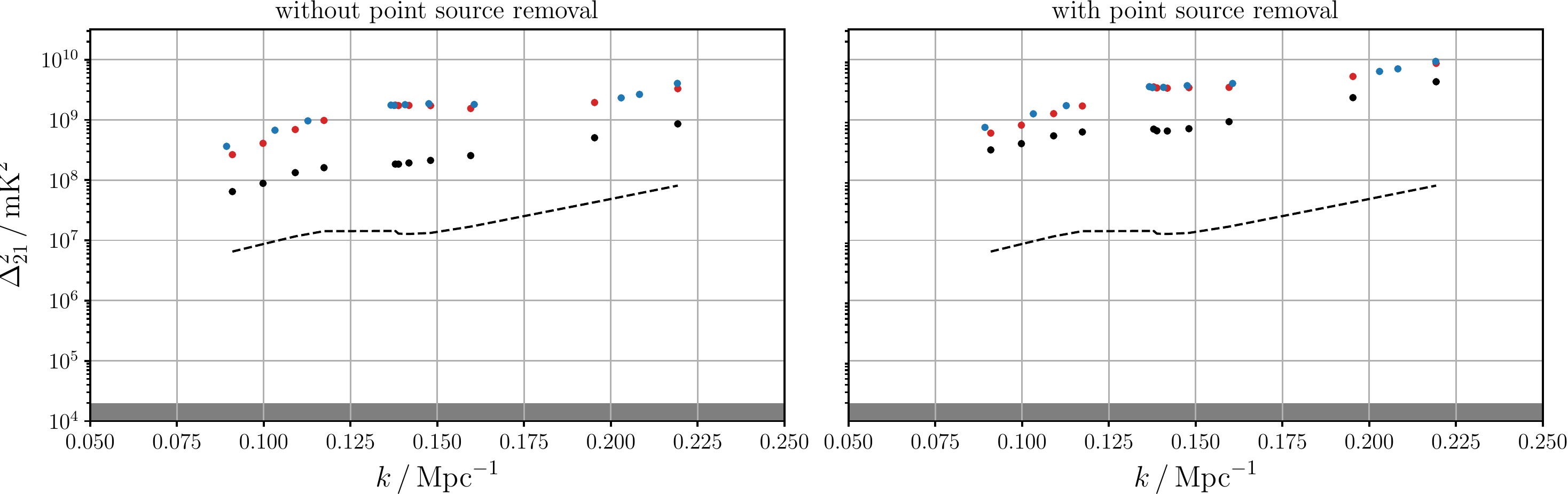}\\
    \end{tabular}
    \caption{
        (top) Mollweide projection of the sky in galactic coordinates after differencing even and
        odd-numbered integrations. The Sun is the dominant artifact in this image due to the
        sporadic failure of source subtraction. Large residuals are also present at low declinations
        that do not rise above $10\arcdeg$ elevation. These low-elevation artifacts are generated by
        RFI.
        (bottom) The power spectrum estimated without point source removal (left) and with point
        source removal (right). Point correspond to the estimated power spectrum amplitude and the
        dashed line corresponds to the computed thermal noise (95\% confidence). Red and blue points
        are estimates from the even and odd numbered integrations respectively. Black points are
        estimates after computing the difference between the two halves of data.
        \revision{The shaded region represents the sensitivity
        required to detect or rule out optimistic models for the 21\,cm power spectrum.}
    }
    \label{fig:spherical-power-spectra-even-odd}
\end{figure*}

Errors arising from variations on rapid timescales---the timescale of a single correlator dump---can
be revealed through the comparison of results obtained data using only even-numbered integrations
and the interleaving odd-numbered integrations. These two halves of the dataset have independent
thermal noise with additional errors due to ionospheric scintillation, RFI and source subtraction
errors.

In prior work we observed that ionospheric scintillation generates $\sim10\%$ fluctuations in the
flux of a point source on 13\,s timescales at 73\,MHz \citep{2018AJ....156...32E}. The position of a
source varies more slowly by up to $4\arcmin$ on 10\,min timescales. Therefore comparing even and
odd-numbered integrations will reveal errors arising from ionospheric scintillation, but not
necessarily from variable ionospheric refraction.

Figure~\ref{fig:spherical-power-spectra-even-odd} contains a map of the sky constructed from
differencing the even and odd datasets (after point source removal).  This map is almost
featureless. If ionospheric scintillation was contributing a substantial amount of additional noise
to the measurement, we would expect to see enhanced residuals in the vicinity of bright point
sources. Instead, the dominant features are a $\sim50\,\text{K}$ residual at the location of the
Sun, and some artifacts that manifest at low declinations that do not rise above $10\arcdeg$
elevation (likely generated by RFI). We therefore conclude that over a long 28\,hr integration, the
ionospheric scintillation has averaged down and is not the dominant source of error.

The bottom panel of Figure~\ref{fig:spherical-power-spectra-even-odd} compares the amplitude of the
estimated power spectrum after differencing the even and odd-numbered integrations. Differencing the
two halves of the dataset cancels out the majority of the residual contamination of foreground
emission into the measurement. Therefore the power spectrum decreases in amplitude. The improvement
is roughly one order of magnitude before source subtraction and only a factor of 2--3 after source
subtraction. Point source removal is conducted independently on each integration, sporadic errors
and source subtraction residuals will therefore also tend to manifest on the timescale of a single
integration. This measurement therefore suggests that source subtraction residuals could be a
limiting factor for this estimate of the 21\,cm power spectrum.

\subsection{Day--Night Jackknife}

\begin{figure*}[t]
    \centering
    \begin{tabular}{c}
        \includegraphics[width=\textwidth]{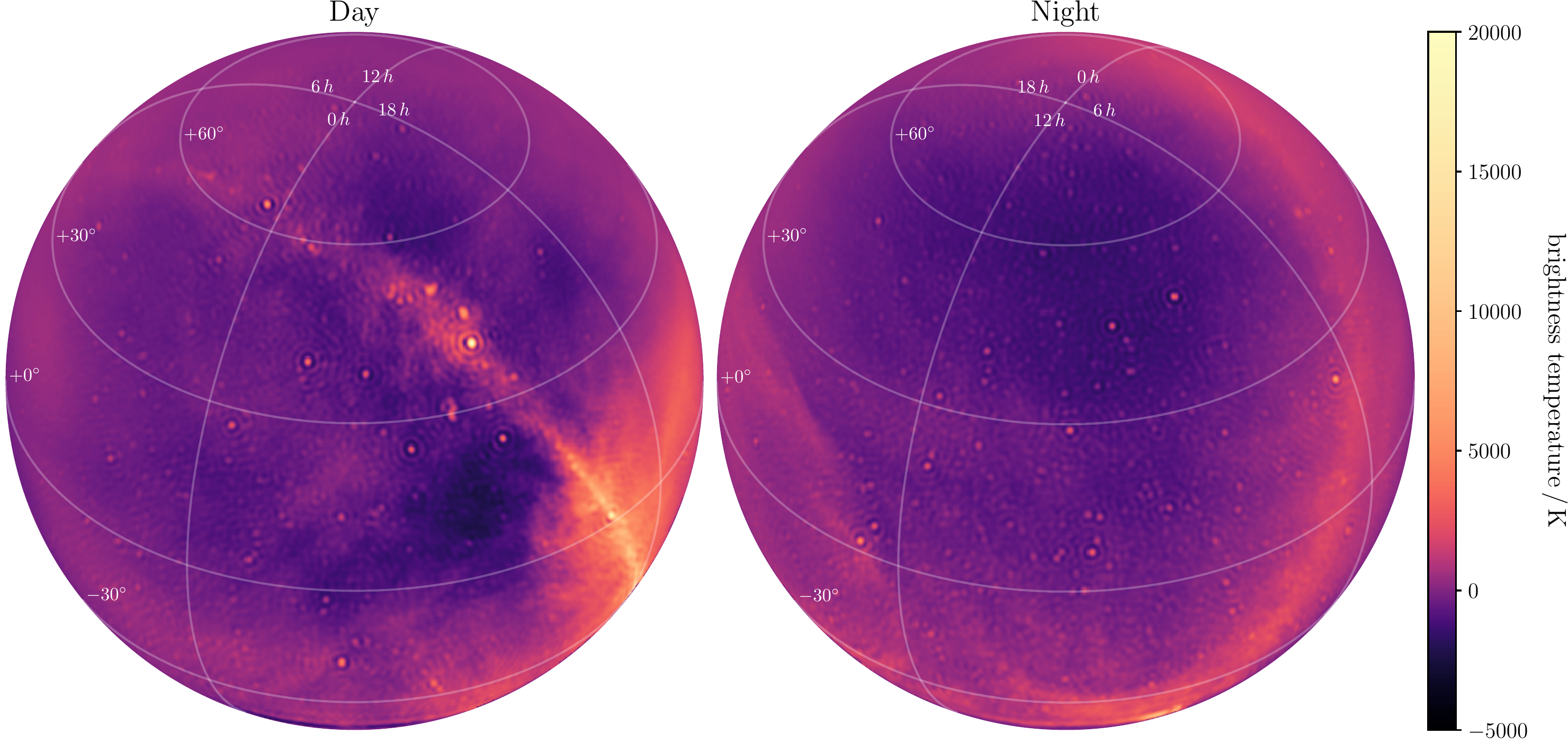}\\
        \includegraphics[width=\textwidth]{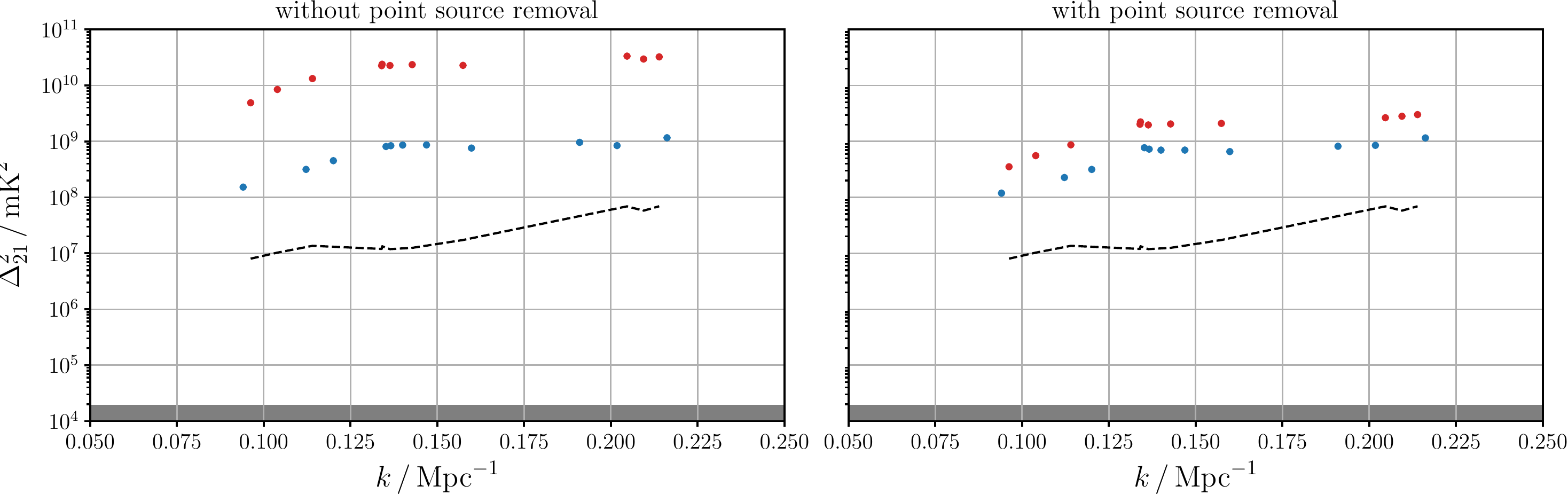}\\
    \end{tabular}
    \caption{
        (top) Orthographic projection of the sky constructed from data collected only during the day
        (left), and only during the night (right).
        (bottom) The power spectrum estimated without point source removal (left) and with point
        source removal (right). Points correspond to the estimated power spectrum amplitude and the
        dashed line corresponds to the computed thermal noise (95\% confidence). Measurements from
        the day are red, and measurements from the night are blue.
        \revision{The shaded region represents the sensitivity
        required to detect or rule out optimistic models for the 21\,cm power spectrum.}
    }
    \label{fig:spherical-power-spectra-day-night}
\end{figure*}

The dominant subtraction residual in the preceding section is associated with the Sun, which is a
difficult source to cleanly subtract due to its complex structure. We can therefore split the data
into two halves: data collected while the Sun is above the horizon, and below the horizon. The data
collected during the night has a number of advantages. Specifically, subtraction residuals
associated with the Sun cannot impact data collected during the night.  Additionally the ionospheric
Total Electron Content (TEC) is lower during the night because the Sun acts as a source of
ionization for the ionosphere. Specifically, the median vertical TEC measured within 200\,km of OVRO
rose to 20\,TECU during the day, but drops to 6\,TECU during the night. There were no geomagnetic
storms during the observing period and the fact that these observations were collected during the
winter months generally contributes to a reduction in the ionospheric TEC.  Finally, due to the time
of year, the sky temperature is lower at night. For these reasons, we generally expect an
improvement in the nighttime data with respect to the daytime data.

In principle, $m$-mode analysis requires that data be collected for a full sidereal day because the
$m$-modes are computed from the Fourier transform of the visibilities with respect to sidereal time.
We relax that requirement here.  When selecting half the data, we additionally apply a
Blackman--Harris window function to prevent ringing. Tikhonov-regularized images made from just the
daytime and nighttime data can be seen in Figure~\ref{fig:spherical-power-spectra-day-night}. These
dirty images serve as a proof of concept that $m$-mode analysis can reasonably be applied to
datasets without a full sidereal day's worth of data.

We estimated the power spectrum from each half of data and the results are presented in
Figure~\ref{fig:spherical-power-spectra-day-night}. Restricting the observations to nighttime-only
leads to a substantial improvement in the power spectrum limits both with and without point source
removal. In fact the measurements with and without point source removal are now comparable. This
suggests the point source subtraction residuals are less of an issue in the nighttime data due to
the fact that (due to the time of year) there are fewer bright point sources that were removed.

\subsection{$xx$--$yy$ Jackknife}

\begin{figure*}
    \centering
    \begin{tabular}{c}
        \includegraphics[width=\textwidth]{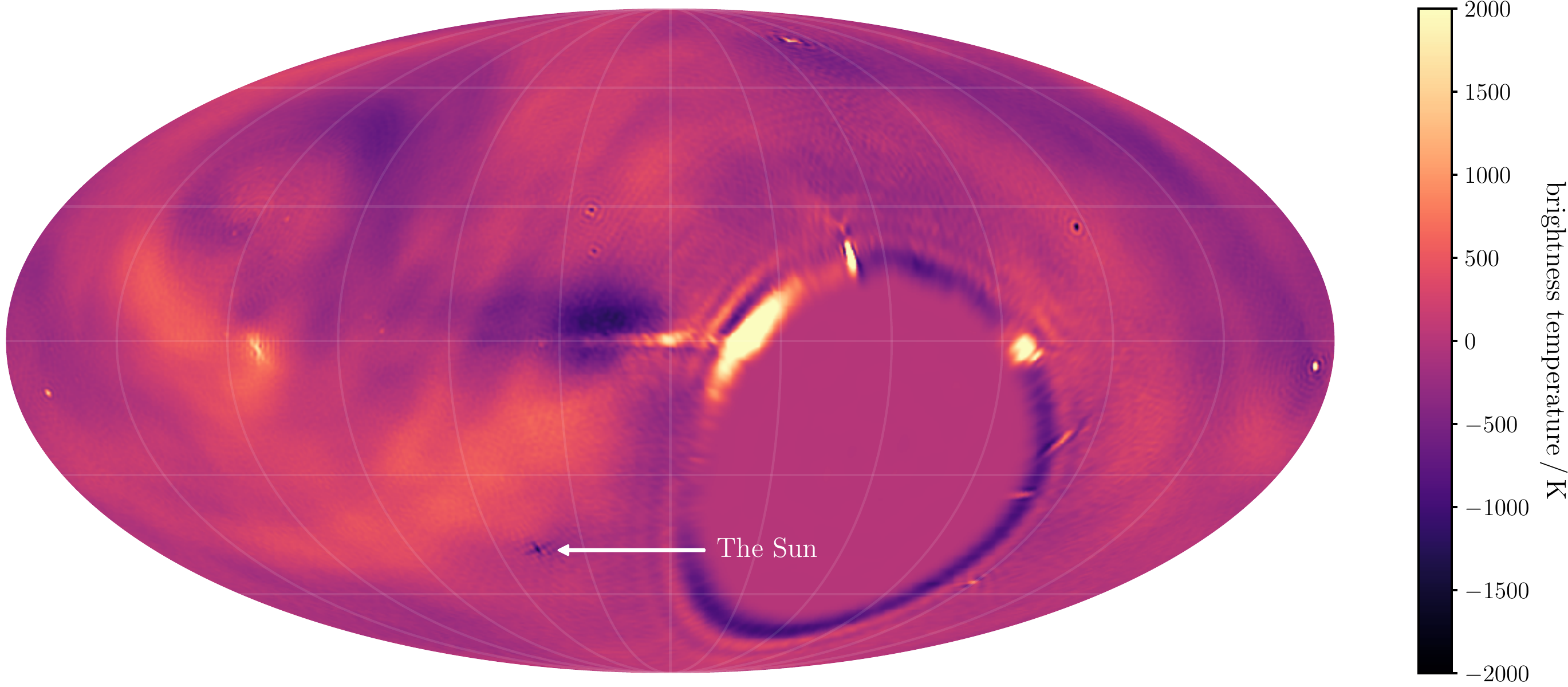}\\
        \includegraphics[width=\textwidth]{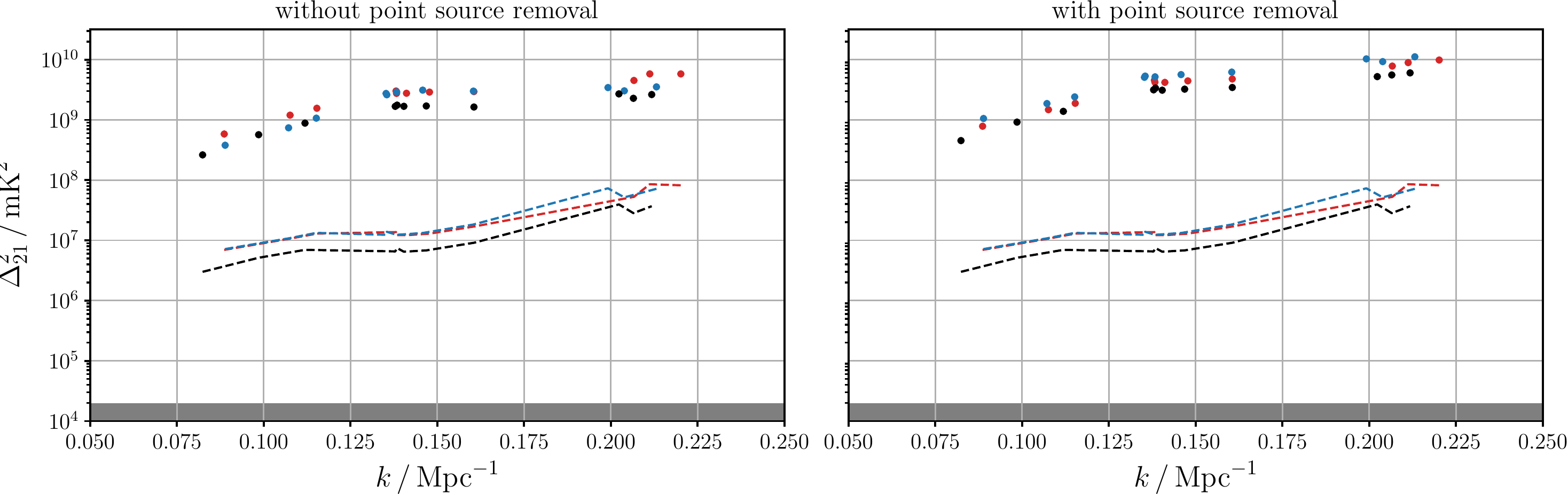}\\
    \end{tabular}
    \caption{
        (top) Mollweide projection of the sky in galactic coordinates after differencing the $xx$
        and $yy$ correlations. Note that this is not true linear polarization because it does not
        account for the full polarization of the antenna response pattern.
        (bottom) The power spectrum estimated without point source removal (left) and with point
        source removal (right). Point correspond to the estimated power spectrum amplitude and the
        dashed line corresponds to the computed thermal noise (95\% confidence). Red and blue points
        are estimates from the $xx$ and $yy$ correlations respectively. Black points are estimates
        from the mean of the $xx$ and $yy$ correlations.
        \revision{The shaded region represents the sensitivity
        required to detect or rule out optimistic models for the 21\,cm power spectrum.}
    }
    \label{fig:spherical-power-spectrum-xx-yy}
\end{figure*}

The polarization angle of linearly polarized emission rotates as it propagates through a magnetized
plasma \citep[e.g.,][]{2014A&A...568A.101J}. The rotation angle is $\propto\lambda^2$, where
$\lambda$ is the wavelength of the radiation. Therefore instrumental polarization leakage from a
linear polarization (Stokes~$Q$ or Stokes~$U$) into total intensity (Stokes~$I$) can introduce
additional spectral structure into the foreground emission that is not accounted for in our
currently unpolarized analysis.

If Faraday-rotated linearly polarized emission is a problem, it will be exacerbated by computing the
power spectrum from the $xx$ correlations and $yy$ correlations separately. For this comparison, the
transfer matrix $\b B$ must be recomputed using the correct response pattern for the individual
dipoles. An image of the sky computed from the difference of the $xx$ and $yy$ correlations is shown
in Figure~\ref{fig:spherical-power-spectrum-xx-yy}. This map is related to the linear polarization
of the sky, but does not account for the full polarization of the beam, and therefore includes some
amount of instrumental polarization.

We estimated the 21\,cm power spectrum from the $xx$ and $yy$ correlations. This estimate is shown
in Figure~\ref{fig:spherical-power-spectrum-xx-yy}. The estimates are comparable to the total
intensity estimate and we therefore conclude that polarization leakage is not currently a major
source of systematic error.

\subsection{Calibration Errors}\label{sec:calibration-errors}

\begin{figure*}
    \centering
    \begin{tabular}{c}
        \includegraphics[width=\textwidth]{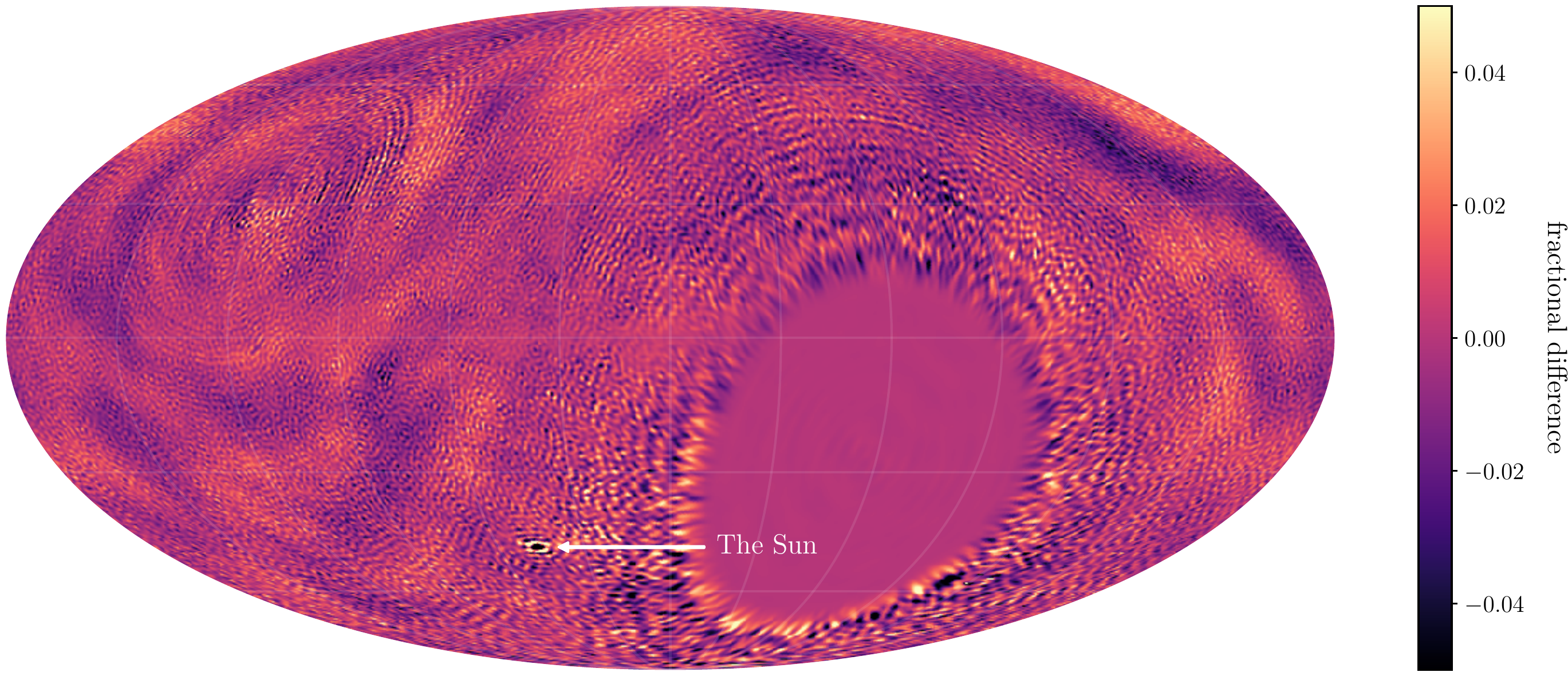}\\
        \includegraphics[width=\textwidth]{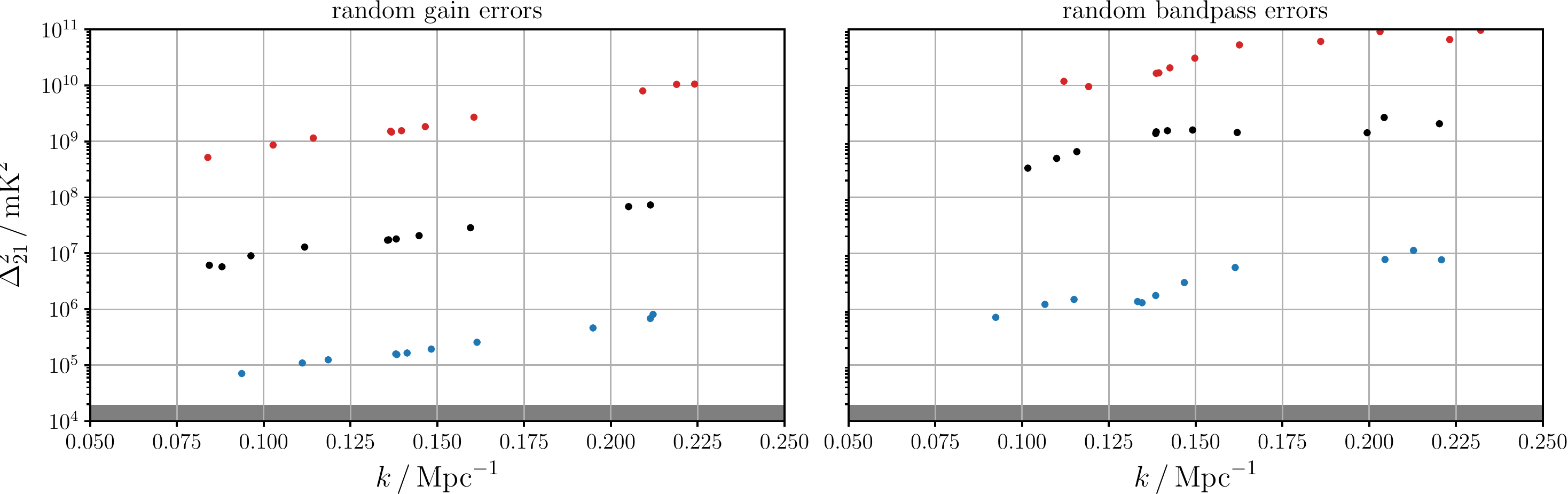}\\
    \end{tabular}
    \caption{
        (top) Mollweide projection of the sky in galactic coordinates after differencing two
        adjacent 240\,kHz frequency channels.
        (bottom) Simulated power spectrum estimates as a result of a foreground model and gain
        errors that are incoherent between antennas (left) and coherent between antennas (right).
        Blue corresponds to 0.1\% errors, black corresponds to 1\% errors, and red corresponds to
        10\% errors in the complex gains.
        \revision{The shaded region represents the sensitivity
        required to detect or rule out optimistic models for the 21\,cm power spectrum.}
    }
    \label{fig:spherical-power-spectrum-gain-errors}
\end{figure*}

Multiple authors have investigated the impact of calibration errors on an experiment's ability to
separate foreground emission from the cosmological 21\,cm signal \citep{2016MNRAS.461.3135B,
2017MNRAS.470.1849E}. In this section we will compute the impact of calibration errors on the double
KL transform foreground filter.

In this calculation will simulate a realistic set of visibilities for the foreground emission, and
introduce errors into the calibration before applying the double KL transform filter. Finally we
will estimate the power spectrum amplitude as a way to characterize the amount of contamination
associated with the calibration errors.

The angular structure of the foreground model used here is measured from the data itself (shown in
the bottom panel of Figure~\ref{fig:before-after-source-removal-sky-maps}), but the frequency
dependence of this emission is chosen to be a power-law with a fiducial spectral index of $-2.3$.
This spectral index was chosen to be consistent with the results reported by LEDA
\citep{2018MNRAS.478.4193P}, but due to the small fractional bandwidth of this measurement, we
expect these results to be insensitive to the specific choice of spectral index. The set of
$m$-modes we expect to measure with the interferometer $\b v_\text{simulated}$ is computed:
\begin{equation}
    \b v_\text{simulated} = \b B\b a_\text{simulated}\,,
\end{equation}
where $\b B$ is the transfer matrix, and $\b a_\text{simulated}$ is a vector of the spherical
harmonic coefficients of the foreground model.

At this point the simulated $m$-modes are corrupted with calibration errors. We explore two
possibilities:
\begin{enumerate}
    \item each antenna and frequency channel receives an incorrect gain calibration
    \item each frequency channel receives an incorrect gain calibration, but this error is coherent
        across antennas
\end{enumerate}
In each case, the gain errors are drawn from a complex normal distribution, and the amplitude of the
error is varied between 0.1\%, 1\%, and 10\%. The former case is coined ``random gain errors'' to
indicate that each antenna is given an error in its complex gain calibration. The latter case is
coined ``random bandpass errors'' to indicate that the overall bandpass of the interferometer is
perturbed. The impact of these calibration errors can be seen in
Figure~\ref{fig:spherical-power-spectrum-gain-errors}.

In order to avoid biasing the 21\,cm power spectrum, these results indicate that the gain
calibration must be derived to an accuracy better than 0.1\%.  A general rule of thumb for the
OVRO-LWA is that for equal amplitude errors, the foreground contamination generated by bandpass
errors that are coherent across all antennas are an order of magnitude worse than for the random
gain errors (in units of $\Delta_{21}^2$). Therefore to achieve a comparable level of foreground
contamination, the overall bandpass of the interferometer must be known to better than 0.01\%.

The dataset presented in this paper is systematically limited at roughly $\Delta_{21}^2 \sim
(10^4\,\text{mK})^2$. These limits are therefore consistent with $\sim 1\%$ errors in the overall
bandpass of the interferometer. The top panel of
Figure~\ref{fig:spherical-power-spectrum-gain-errors} therefore presents the fractional difference
in the sky images between two adjacent frequency channels (after averaging down to 240\,kHz channel
resolution). The residuals in this sky map are typically 2\%--3\%, but generally do not correlate
with the sky brightness. We therefore conclude that between adjacent 240\,kHz channels, the bandpass
error is less than 1\%.

In fact, the structure of the residuals in Figure~\ref{fig:spherical-power-spectrum-gain-errors}
suggests a different terrestrial source. Terrestrial sources of radio emission do not move through
the sky at a sidereal rate. Therefore when constructing images of the sky, this contaminating
emission tends to be smeared along rings of constant declination. These ring-like structures are
clearly visible in Figure~\ref{fig:spherical-power-spectrum-gain-errors} alongside some larger scale
diffuse structures.

However, if we attribute the residual emission entirely to gain errors, then these simulations
suggest that the antenna gains are known only to within a couple percent, which when considered
alongside the RFI that contaminates the measurement, is likely sufficient to explain the current
systematic limitations of our dataset.

\revision{
    The results of this section additionally put some constraints on the impact of errors in the
    foreground covariance computed in \S\ref{sec:foreground-covariance}. For instance, a $\sim0.1$
    error in the spectral index of the foreground emission may be interpreted as a $\sim0.5\%$
    spectrally smooth bandpass error that will degrade the performance of the foreground filter.
    However, because these errors are spectrally smooth, the performance of the foreground filter
    will degrade less than due to the random bandpass errors simulated in this section. We therefore
    interpret the calculations performed here as an upper bound. A $\sim0.1$ error in the spectral
    index of the foreground emission will lead to systematic errors less than a comparable
    $\sim0.5\%$ spectrally unsmooth error in the bandpass, which may be inferred from
    Figure~\ref{fig:spherical-power-spectrum-gain-errors}.
}

\subsection{Non-Stationarity}\label{sec:non-stationary}

\revision{
    One implicit assumption of $m$-mode analysis is that all measured components are stationary: the
    foreground emission, the 21\,cm signal, and the thermal noise.
}

\revision{
    For foreground components, we assumed that the emission is a Gaussian random field.
    \citet{2014ApJ...781...57S, 2015PhRvD..91h3514S} demonstrated that under ideal circumstances
    this assumption on its own does not substantially bias CHIME estimates of the 21\,cm power
    spectrum.  However, when the antenna beam is not perfectly measured, foreground emission leaks
    through the foreground filters to corrupt the measurement. We also see this in
    \S\ref{sec:calibration-errors}, where bandpass errors lead to residual foreground emission
    contaminating the measurement, but in the absence of bandpass errors the real non-Gaussian
    foreground emission does not directly limit the measurements.
}

\revision{
    An additional source of non-stationarity can be seen in Figure~\ref{fig:Tsys}, which shows that
    the system temperature varies by approximately 50\% over the course of a sidereal day. We
    simulated the impact of this variation in system temperature by computing visibilities from a
    realistic sky model (see Figure~\ref{fig:before-after-source-removal-sky-maps}) and injecting
    noise into the measured visibilities. No 21\,cm signal is injected into the measurement because
    we're simply trying to measure the impact of the injected noise on foreground filtering and
    power spectrum estimation. We ran the simulation twice. In the first case, the system
    temperature varied sidereally and was directly sampled from the data presented in
    Figure~\ref{fig:Tsys}. In the second case, the system temperature was held fixed at the mean
    system temperature. After foreground filtering and power spectrum estimation, the two estimates
    were indistinguishable within the computed uncertainties. We therefore conclude that sidereal
    variation of the system temperature is not a limiting factor in our experiment.
}

\section{Conclusion}\label{sec:conclusion}

In this paper we estimated the amplitude of the 21\,cm power spectrum of the Cosmic Dawn with 28\,hr
of data from the OVRO-LWA. This measurement was severely limited by systematic errors and therefore
we interpret our measurements as upper limits, which are currently the most sensitive at this epoch
and the first measurement at $z > 18$. We measured $\Delta_{21}^2 \lesssim (10^4\,\text{mK})^2$ at
$k \approx 0.10\,\text{Mpc}^{-1}$.

In making this measurement, we demonstrated the first application of the double KL transform
foreground filter to a measured dataset. We demonstrated that the application of this foreground
filter can lead to improved power spectrum limits, and in combination with Tikhonov-regularized
imaging, we developed a physical intuition for the action of the foreground filter.  The double KL
transform derives its action from models for the foreground and 21\,cm signal covariance. We
measured the angular power spectrum of the foreground emission and found that the power-law index
appears to steepen on large angular scales ($l < 50$). The 21\,cm signal covariance is derived from
the flat-sky approximation, which we derive in Appendix~\ref{app:spatial-to-angular}.

Although application of the foreground filter leads to some improvement in our measurement of the
21\,cm power spectrum, the improvement was relatively modest. This is essentially a reflection of
the fact that the true covariance of the data does not match the expectations of the models. We
performed a series of jackknife tests and simulations that appear to implicate a combination of
source subtraction errors, terrestrial interference, and calibration errors as limiting factors in
this measurement.

\revision{
    If this measurement was thermal noise limited, we may have expected to place limits at the level
    of $\Delta_{21}^2\sim (10^3\,\text{mK})^2$ to $(10^{3.5}\,\text{mK})$ depending strongly on the
    amount of foreground filtering that is needed to adequately suppress the foreground
    contamination.  Because the action of the foreground filter is to exchange thermal noise
    sensitivity for more comprehensive removal of foreground emission, extrapolating from here to
    find the requisite sensitivity of an interferometer that will be able to make a detection is
    difficult.  An improved calibration may greatly reduce the required integration time, while an
    increased need for foreground filtering may extend the required integration time.  However, from
    simple scaling arguments, and in an optimistic scenario where the experiment does attain its
    noise limited sensitivity, the OVRO-LWA could begin to constrain the brightest models of the
    21\,cm power spectrum ($\Delta_{21}^2\sim(10^2\,\text{mK})^2$) with between $10^3$ and $10^4$
    hours of observing time.
}
However, a detection of the Cosmic Dawn 21\,cm spatial power spectrum will require that gain errors
are restricted to less than 0.1\% and bandpass errors to less than 0.01\%.  Improved results will
certainly require improving the instrumental calibration and source removal, which will help to
prevent foreground emission leaking through the measurement and into the power spectrum estimate.

\acknowledgments

This material is based in part upon work supported by the National Science Foundation under grants
AST-1654815 and AST-1212226. The OVRO-LWA project was initiated through the kind donation of Deborah
Castleman and Harold Rosen.

Part of this research was carried out at the Jet Propulsion Laboratory, California Institute of
Technology, under a contract with the National Aeronautics and Space Administration, including
partial funding through the President's and Director's Fund Program.

This work has benefited from open-source technology shared by the Collaboration for Astronomy Signal
Processing and Electronics Research (CASPER).  We thank the Xilinx University Program for donations;
NVIDIA for proprietary tools, discounts, and donations; and Digicom for collaboration on the
manufacture and testing of DSP processors.

We thank the Smithsonian Astrophysical Observatory Submillimeter Receiver Lab for the collaboration
of its members, and the observatory for research and development funds.

Development, adaptation, and operation of the LEDA real-time digital signal-processing systems at
OVRO-LWA have been supported in part by NSF grants AST/1106059, PHY/0835713, OIA/1125087, and
AST/1616709.

\appendix

\section{Converting a Spatial Power Spectrum to an Angular Power Spectrum}
\label{app:spatial-to-angular}

The multi-frequency angular power spectrum $C_l(\nu, \nu^\prime)$ is measured from the spherical
harmonic coefficients of the sky $a_{lm}(\nu)$ at the frequencies $\nu$ and $\nu^\prime$:
\begin{equation}
    C^{21}_l(\nu, \nu^\prime) = \frac{1}{2l+1}\sum_{m = -l}^l \left\langle
        a^{21}_{lm}(\nu) \, {a^{21}_{lm}}^*(\nu^\prime)
    \right\rangle\,,
\end{equation}
where the angled brackets should be understood as an ensemble average over sky realizations. Here
the average over $m$ is indicated with an explicit sum to distinguish it from the ensemble average.

The spherical harmonic coefficients themselves are computed from an integral over the 21\,cm
brightness temperature $T^{21}_\nu(\vec r)$ over a spherical shell of the universe.
\begin{equation}
    a^{21}_{lm}(\nu) = \int T^{21}_\nu(\vec r) \, Y_{lm}^*(\hat r) \, \delta(r - r_z) \, \d^3r\,,
\end{equation}
where $Y_{lm}(\hat r)$ is a spherical harmonic function, and the Dirac delta function
$\delta(r-r_z)$ is used to pick out the spherical shell of the universe at the comoving distance
$r_z$ to the redshift $z$.

The 21\,cm brightness temperature is related to the power spectrum $P^{21}_z(\vec k)$ through its
Fourier transform.
\begin{align}
    T^{21}_\nu(\vec r) &=
        \int T^{21}_\nu(\vec k) \, e^{i\vec k\cdot\vec r} \, \frac{\d^3k}{(2\pi)^3} \\
    \left\langle T_\nu(\vec k) T_{\nu^\prime}(\vec k^\prime) \right\rangle &=
        (2\pi)^3 \, \delta^3(\vec k - \vec k^\prime) \, P^{21}_z(\vec k)
\end{align}

Finally, we will need the ``plane wave expansion'' that describes a plane wave in terms of spherical
harmonics:
\begin{equation}
    e^{i\vec k\cdot\vec r} = 4\pi \sum_{lm} i^l \, j_l(kr) \, Y_{lm}(\hat r) \, Y^*_{lm}(\hat k)\,,
\end{equation}
where the function $j_l(kr)$ is the spherical Bessel function of the first kind.

Putting this all together we can find
\begin{align*}
    C^{21}_l(\nu, \nu^\prime) &=
        \frac{1}{2l+1}\sum_{m=-l}^l
        \left\langle
            \iiiint
            T_\nu^{21}(\vec k) {T_{\nu^\prime}^{21}}^*(\vec k^\prime) \,
            e^{i(\vec k\cdot\vec r - \vec k^\prime\cdot\vec r^\prime)} \,
            Y_{lm}^*(\hat r) \,
            Y_{lm}(\hat r^\prime) \,
            \delta(r - r_z) \,
            \delta(r^\prime - r_{z^\prime}) \,
            \d^3r \,
            \d^3r^\prime \,
            \frac{\d^3k}{(2\pi)^3} \,
            \frac{\d^3k^\prime}{(2\pi)^3}
        \right\rangle \\
    &=
        \frac{4\pi(-i)^l}{2l+1}\sum_{m=-l}^l
        \left\langle
            \iiint
            T_\nu^{21}(\vec k) {T_{\nu^\prime}^{21}}^*(\vec k^\prime) \,
            e^{i\vec k\cdot\vec r} \,
            j_l(k^\prime r_{z^\prime}) \,
            Y_{lm}^*(\hat r) \,
            Y_{lm}(\hat k^\prime) \,
            \delta(r - r_z) \,
            \d^3r \,
            \frac{\d^3k}{(2\pi)^3} \,
            \frac{\d^3k^\prime}{(2\pi)^3}
        \right\rangle \\
    &=
        \frac{(4\pi)^2}{2l+1}\sum_{m=-l}^l
        \left\langle
            \iint
            T_\nu^{21}(\vec k) {T_{\nu^\prime}^{21}}^*(\vec k^\prime) \,
            j_l(k r_z) \,
            j_l(k^\prime r_{z^\prime}) \,
            Y_{lm}^*(\hat k) \,
            Y_{lm}(\hat k^\prime) \,
            \frac{\d^3k}{(2\pi)^3} \,
            \frac{\d^3k^\prime}{(2\pi)^3}
        \right\rangle \\
    &=
        \frac{(4\pi)^2}{2l+1}\sum_{m=-l}^l
        \int
        P_z^{21}(\vec k) \,
        j_l(k r_z) \,
        j_l(k r_{z^\prime}) \,
        Y_{lm}^*(\hat k) \,
        Y_{lm}(\hat k) \,
        \frac{\d^3k}{(2\pi)^3} \\
\end{align*}
Where in the first two steps we used the plane-wave expansion, and in the final step we used the
definition of the spatial power spectrum.

\begin{figure*}
    \centering
    \includegraphics[width=\textwidth]{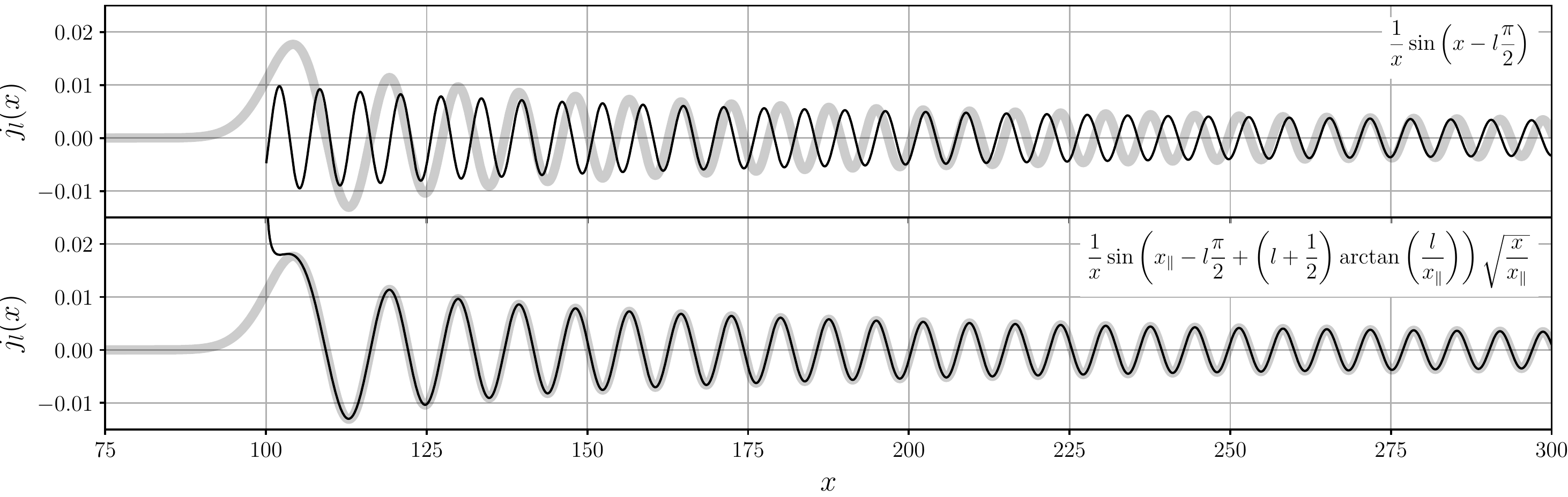}
    \caption{
        A comparison of two approximations (thin black lines) to the spherical Bessel function
        $j_l(x)$ with $l=100$ (thick gray lines). The top panel shows the approximation derived from
        the limiting behavior of $j_l(x)$ as $x\rightarrow\infty$. This is a poor approximation near
        $x\sim l$. The bottom panel shows the approximation derived from the method of steepest
        descent. This approximation maintains the same limiting behavior as $x\rightarrow\infty$ and
        greatly improves the accuracy of the approximation near $x\sim l$.
    }
    \label{fig:bessel-approximations}
\end{figure*}

At this point if we assume that the power spectrum is isotropic and has no dependence on the
orientation of the wave vector $\hat k$, then the angular component of the remaining integral can be
performed to find:
\begin{equation}\label{eq:csignal-basic}
    C^{21}_l(\nu, \nu^\prime) =
        \frac{2}{\pi}
        \int
        P_z^{21}(k) \,
        j_l(k r_z) \,
        j_l(k r_{z^\prime}) \,
        k^2 \, \d k\,.
\end{equation}
Typically $r_z\sim 10,000\,\text{Mpc}$ and $k\sim 0.1\,\text{Mpc}^{-1}$ so the spherical Bessel
functions $j_l(x)$ are typically evaluated in the limit of $l < x < l^2$.
Equation~\ref{eq:csignal-basic} is exact for an isotropic power spectrum, but in practice integrals
over the product of two spherical Bessel functions are numerically challenging due to their
oscillatory behavior, so we will look for a scheme to approximate this integral.  This approximation
is simple in the regimes where $x \ll l$ and $x \gg l$, because
\begin{align}
    \lim_{x\rightarrow 0} j_l(x) &\propto x^l \approx 0 \\
    \lim_{x\rightarrow\infty} j_l(x) &=
        \frac{1}{x} \sin\left(x - l\frac{\pi}{2}\right)
        + \mathcal{O}\left(\frac{l^2}{x^2}\right)
    \,.
\end{align}
However, we are primarily interested in the intermediate regime where a better approximation can be
obtained using the method of steepest descent (this method is also used, for example, to derive
Stirling's approximation to $\log n!$ for large $n$). Starting with the integral representation of
the spherical Bessel functions, the integration contour is deformed slightly to pass through saddle
points of the integrand, approaching along paths of steepest descent. This allows the integral to be
approximated as a Gaussian integral, which can be analytically evaluated to
\begin{equation}
    j_l(x) \approx
        \frac{1}{x}
        \sin\left(
            x_\parallel - l\frac{\pi}{2}
            + \left(l+\frac{1}{2}\right)\arctan\left(\frac{l}{x_\parallel}\right)
        \right)
        \sqrt{\frac{x}{x_\parallel}}
    \,,
\end{equation}
where we have suggestively defined $x_\parallel = \sqrt{x^2-l^2}$.  This approximation holds for
$l\gg 1$ and $x > l$ (see Figure~\ref{fig:bessel-approximations}). The product of the two spherical
Bessel functions in Equation~\ref{eq:csignal-basic} therefore results in rapid oscillations on top
of a slower beat frequency. After computing the integral---provided the power spectrum is
sufficiently smooth---the rapid oscillations will average down. Therefore this integral can be
approximated using only the latter term:
\begin{equation}\label{eq:csignal-beats}
    C^{21}_l(\nu, \nu^\prime) \approx
        \frac{1}{\pi r_z r_{z^\prime}}
        \int
        P_z^{21}(k_\perp, k_\parallel) \,
        \cos\left(
            k_\parallel \Delta r_z
            + \left(l+\frac{1}{2}\right) \arctan\left(
                \frac{\Delta r_z/r_z}{k_\perp/k_\parallel+k_\parallel/k_\perp}
            \right)
        \right)
        \, \d k_\parallel\,,
\end{equation}
where $k_\parallel = \sqrt{k^2 - k_\perp^2}$, $k_\perp = l/r_z$, and $\Delta r_z = r_{z^\prime} -
r_z$. Typically for nearby frequency channels, the argument to the arctangent will be small, and
consequently we arrive at the ``flat-sky approximation'' used by \citet{2005MNRAS.356.1519B} and
\citet{2007MNRAS.378..119D}:
\begin{equation}
    C^{21}_l(\nu, \nu^\prime) \approx
        \frac{1}{\pi r_z r_{z^\prime}}
        \int
        P_z^{21}(k_\perp, k_\parallel) \,
        \cos\left(k_\parallel \Delta r_z\right)
        \, \d k_\parallel\,.
\end{equation}
In order to derive this approximation we required $l\gg 1$, $\Delta r_z/r_z \ll
k_\parallel/k_\perp$, and that the power spectrum is smooth on scales that allow the rapid
oscillations of the spherical Bessel functions to average down.  The flat-sky approximation is
advantageous because piecewise linear representations of the power spectrum can be rapidly evaluated
analytically. However, we separately need to verify that the piecewise linear representation is
smooth enough to permit this approximation. \revision{This assumption is evaluated in
\S\ref{sec:signal-covariance} and Figure~\ref{fig:flat-sky-approximation} for a fiducial piecewise
linear power spectrum.}

\bibliographystyle{aasjournal}
\bibliography{paper}

\end{document}